\documentclass[useAMS,usenatbib,usegraphicx,mathbb]{mn2e}
\title[LMC Symbiotic Stars in OGLE Data]{Symbiotic Stars in OGLE Data \\ I. Large Magellanic Cloud Systems}
\author[R. Angeloni et al.]
  {R.~Angeloni$^{1,2,3}$,
  C.~E.~Ferreira Lopes$^{4,1,3}$,
  N.~Masetti$^5$,
  F.~Di Mille$^{6}$,
  P.~Pietrukowicz$^7$,\newauthor
  A.~Udalski$^7$,
  B.~E.~Schaefer$^{8}$,
  P.~Parisi$^{9}$,
  R.~Landi$^5$,
  C.~Navarrete$^{1,3}$,
  M.~Catelan$^{1,3}$, \newauthor
  T.~H.~Puzia$^1$\& 
  D.~Guzman$^2$
  \\
  $^1$Instituto de Astrof\'isica, Pontificia Universidad Cat\'olica de Chile, Vicu\~{n}a Mackenna 4860, 7820436 Macul, Santiago, Chile\\
  $^2$Departamento de Ingenier\'ia El\'ectrica, Pontificia Universidad Cat\'olica de Chile, Vicu\~{n}a Mackenna 4860, 7820436 Macul, Santiago, Chile\\
  $^3$The Milky Way Millennium Nucleus, Vicu\~{n}a Mackenna 4860, 7820436 Macul, Santiago, Chile\\
  $^4$Universidade Federal do Rio Grande do Norte,  Campus Universit\'ario Lagoa Nova CEP 59078-970, Natal, Brasil\\
  $^5$INAF - Istituto di Astrofisica Spaziale e Fisica Cosmica di Bologna, via Gobetti 101, 40129, Bologna, Italy\\
  $^6$Dipartimento di Fisica e Astronomia, Universit\`{a} degli Studi di Padova, Vicolo dell' Osservatorio 3, 35122 Padova, Italy\\
  $^7$Warsaw University Astronomical Observatory, Al. Ujazdowskie 4, 00-478, Warszawa, Poland\\  
  $^8$Department of Physics and Astronomy, Louisiana State University, Baton Rouge, LA 70803-4001, USA\\
  $^9$INAF - Istituto di Astrofisica e Planetologia Spaziali, Via Fosso del Cavaliere 100, 00133, Roma, Italy}
  
\date{Released 2002 Xxxxx XX}

\pagerange{\pageref{firstpage}--\pageref{lastpage}} \pubyear{2002}

\def\LaTeX{L\kern-.36em\raise.3ex\hbox{a}\kern-.15em
    T\kern-.1667em\lower.7ex\hbox{E}\kern-.125emX}

\begin{document}

\label{firstpage}

\maketitle

\begin{abstract}
Symbiotic stars are long-orbital-period interacting-binaries characterized by extended emission over the whole electromagnetic range and by complex photometric and spectroscopic variability.  In this paper, the first of a series, we present OGLE light curves of all the confirmed symbiotic stars in the Large Magellanic Cloud, with one exception. By careful visual inspection and combined time-series analysis techniques, we investigate for the first time in a systematic way the photometric properties of these astrophysical objects, trying in particular to distinguish the nature of the cool component (e.g., Semi-Regular Variable vs. OGLE Small-Amplitude Red Giant), to provide its first-order pulsational ephemerides, and to link all this information with the physical parameters of the binary system as a whole. Among the most interesting results, there is the discovery of a 20-year-long steady fading of Sanduleak's star, a peculiar symbiotic star known to produce the largest stellar jet ever discovered.  
We discuss by means of direct examples the crucial need for long-term multi-band observations to get a real understanding of symbiotic and other interacting binary stars.  We eventually introduce BOMBOLO, a multi-band simultaneous imager for the SOAR 4m Telescope, whose design and construction we are currently leading.
\end{abstract}

\begin{keywords}
 binaries: symbiotic - surveys - techniques: photometric - stars: individual: LMC S147, LMC N19, LMC 1, LMC N67, Sanduleak's star, LMC S63, SMP LMC 94.
\end{keywords}

\section{Introduction}
Symbiotic stars (hereafter SySs) are binaries composed of a hot compact star - generally but not necessarily a white dwarf (WD) - and a late type giant, whose mutual interaction through accretion processes triggers the emission recorded from radio to X-rays. Each of the three main emitting components (hot star, cool star, and the network of gas and dust around these stars) dominate the spectral energy distribution (SED) over different wavelength regions (Kenyon 1986). \\

A remarkable aspect of the symbiotic phenomenon is the complex photometric variability patterns caused by a wide range of phenomena.  Some of the mechanisms that lead to large-amplitude variations are the intrinsic stellar pulsations of the giant companion; ellipsoidal variations of the giant; reflection effects due to the strong WD radiation field that is illuminating the facing hemisphere of the giant; mutual eclipses for those systems with a high orbital inclination; dust obscuration events; and thermonuclear outbursts occurring on the surface of the WD.
These effects produce time scales for the variability ranging from flicker in the accretion disk in the second/minute range (e.g., Angeloni et al. 2012a); the rotation of the WDs typically in the minutes/hours range (e.g., as in Z And, Sokoloski \& Bildsten 1999); stellar pulsations of the late giant in the day/months range; and recombination in the nebula in the day/months range, modulations associated with the orbital period in the years range; and nova-like outbursts in the years/century range. \\

Several recent studies have used \textit{Optical Gravitational Lensing Experiment} (OGLE\footnote{http://ogle.astrouw.edu.pl/}) and \textit{Massive Compact Halo Object} (MACHO\footnote{http://macho.anu.edu.au/}) data to study Galactic SySs, including Gromadzki et al. (2009), Lutz et al. (2010), and Miszalski, Mikolajewska \& Udalski (2013).  Mikolajewska (2004) discusses MACHO light curves of 4 Magellanic SySs. Munari (2012) and Skopal (2008) present general reviews with special emphasis on SyS photometric properties.\\

SySs represent unique laboratories for studying a variety of important astrophysical problems.
The energetics is a key ingredient for a meaningful understanding of the physical processes at work, but it relies on accurate knowledge of the distance. Unfortunately, the distances to Galactic SySs have large uncertainties.  This prevents reliable calibration of their energy budget of the interaction processes and the evolutionary status of the companion.  Fortunately, SySs are amongst the intrinsically brightest variable stars, therefore detectable also in nearby galaxies. Out of 18 extragalactic SySs known today, 15 belong to the Magellanic Clouds (hereafter MCs) -- 7 in the Small MC (SMC) and 8 in the Large MC (LMC) -- while 3 have been found in further Local Group galaxies (Draco C-1 -- Munari \& Buson, 1994; IC10 StSy-1 -- Gon\c{c}alves et al., 2008; NGC 6822 SySt–1 -- Kniazev et al., 2009). It is clear that observations of extragalactic SySs do not have the primary ambiguity in studying these interacting stars, namely the lack of reliable distances. \\

SySs in the MCs are interesting in themselves, because they are differentiated from their Galactic cousins in several aspects. Already in the 1970's, SySs were grouped into S- and D-types according to whether the cool star (S-type) or dust (D-type) dominated the near-IR spectral range (Webster \& Allen 1975). In our Galaxy, D-type SySs seem to host invariably a Mira variable, while Red Giant Branch (RGB) stars are generally found in S-types. Like their Galactic cousins, MC SySs contain low mass ($\leq$ 3 M$_{\odot}$) giants as cool components. However, in the MCs, all SySs giants are in the Asymptotic Giant Branch (AGB) phase (M\"{u}rset et al. 1996; Mikolajewska 2004; Kniazev et al. 2009). In the LMC, 4 out of 8 SySs are classified as D-type systems, while only one D-type is known in the SMC (recently discovered by Oliveira et al. 2013). Furthermore, the position in the H-R diagram of the hot component reveals that MC SySs are amongst the hottest and the brightest within the entire 
known 
symbiotic 
population (Mikolajewska 2004). It is likely that at least some of these differences arise because the present MC sample is biased toward the most luminous objects. Further investigations based on new data are needed to understand the differences.\\

As part of a long-term project aimed at characterizing the symbiotic phenomenon outside the Milky Way (e.g. Angeloni et al. 2009, 2011), we are exploring the long-term photometric behaviour of SySs in the MCs. Specifically, in this study we present, for the first time, the OGLE optical light curves for all but one the SySs in LMC (Table \ref{tab:masys} - LMC S154 being the only SyS that is not in the OGLE fields). Whenever available, OGLE-IV data are complemented with data from the third phase of the OGLE Project, and with unpublished archival MACHO Project data. In Section 2, we give the details of the observations (Sect. 2.1 \& 2.2) and of the data analysis general methodology (Sect. 2.3). In Section 3, we describe our findings, presenting each object in individual subsections. A more general discussion of the results follows in Sect. 4, while concluding remarks appear in Sect. 5. In the second paper of the series (Angeloni et al., in preparation), we will complement this study by presenting the OGLE 
monitoring for the whole sample of SySs in the SMC.

\begin{table*}
 \caption{The LMC SySs discussed in the paper.}\label{tab:masys}
 \label{coord}
 \begin{tabular}{@{}lccccccccc}
  \hline
  \#$^a$ & Name$^a$ & SIMBAD & Equat. coords.$^b$ (J2000) & OGLE-III & OGLE-IV \\
        &          &      name       & $\alpha$ $\;$ $\delta$ & field & field \\
  \hline	
  14 & LMC S147 & [BE74] 484 & 04:54:03.43 $\;$ -70:59:32.2 & - & LMC530.02 \\
  15 & LMC N19 & LHA 120-N 19 & 05:03:23.74 $\;$ -67:56:33.5 & LMC125.3 & LMC511.06 \\
  18 & LMC 1 & NAME LMC 1 & 05:25:01.11 $\;$ -62:28:48.9 & - & LMC593.24\\
  19 & LMC N67 & LHA 120-N 67 & 05:36:07.55 $\;$ -64:43:21.3 & - & LMC520.17\\
  20 & Sanduleak's star & NAME Sanduleak's star & 05:45:19.57 $\;$ -71:16:06.7 & LMC186.8 & LMC551.31\\
  21 & LMC S63 & SV* HV 12671 & 05:48:43.42 $\;$ -67:36:10.3 & - & LMC554.26\\
  22 & SMP LMC 94 & 2MASS J05540952-7302341 & 05:54:09.53 $\;$ -73:02:34.1 & - & LMC550.10\\
  \hline
 \end{tabular}
 
 \medskip
 {$a$}: as in the Belczy{\'n}ski et al. (2000) atlas;
 {$b$}: as in the 2MASS Point Source Catalogue (Cutri et al. 2003).
\end{table*}

\section{The Data}

\subsection{OGLE data}
The bulk of the photometric data presented in this paper come from the fourth
and, in some cases, the third phase of the OGLE Project. The project is conducted at the 1.3-m
Warsaw Telescope located at Las Campanas Observatory, Chile,
operated by the Carnegie Institution for Science. Since the
beginning of the fourth phase (OGLE-IV) in March 2010, a 32-chip
mosaic camera is used, covering approximately 1.4 square degrees
on the sky with a scale of 0.26 arcsec/pixel. The third phase
of the project (OGLE-III) was conducted using an 8-chip camera
(with the same scale) during the years 2001-2009.

The main aim of the OGLE survey is searching for gravitational
micro-lensing phenomena in the Galactic bulge (Udalski 2003).
These same images allow for long-term well-sampled photometric monitoring of stars in our Galactic disk,
in the MCs, in the Magellanic Bridge, as well as selected extragalactic
sources such as Huchra's Lens.  OGLE-III observations covered about
40~deg$^2$ in the LMC (Udalski et al. 2008a)
and about 14~deg$^2$ in the SMC (Udalski et al.
2008b). The area monitored in the first two seasons of OGLE-IV
covers about 160 and 55~deg$^2$ over LMC and SMC, respectively.

The OGLE project focuses on variability, so about 90\% of observations
are carried out in the Cousins $I$-band filter. The rest of the images
are taken in the Johnson $V$-band to secure colour information.
OGLE-III light curves for LMC stars usually have $\sim$450 points
in $I$ and $\sim$40 points in $V$. For SMC stars, we typically have $\sim$700 magnitudes
in $I$ and $\sim$50 magnitudes in the $V$. Typically,
OGLE-IV collects $\sim$250 points in $I$ in the densest parts
of the LMC and $\sim$150 points in the SMC per season. Some
MC fields are observed less frequently.

\begin{table*}
 \caption{OGLE observations of LMC SySs.}\label{tab:oglemasys}
 \label{ogle}
 \begin{tabular}{@{}cccccccccccccccc}
  \hline
   Name & \# of epochs & Time span [d] & $<$mag$>$ & Max. ampl.\\
        &  $V$ $\;$ $I$ & $V$ $\;$ $I$ & $V$ $\;$ $I$ & $\Delta V \; \Delta I$\\
  \hline
 LMC S147 & 198 $\;$ 555 & 734 $\;$ 800 & 15.73 $\;$ 14.16 & 0.336 $\;$ 0.194  \\
 LMC N19 & 44 $\;$ 733 & 1898 $\;$ 3892 & 16.83 $\;$ 14.18 & 0.759 $\;$ 0.478\\
 LMC 1 & - $\;$ 87 & - $\;$ 812 & - $\;$ 13.61 & - $\;$ 0.479 \\ 
 LMC N67 &  - $\;$ 287 & - $\;$ 807 & - $\;$ 14.5 & - $\;$ 0.333  \\ 
 Sanduleak's star & 37 $\;$ 738 & 1896 $\;$ 3912 & 17.28 $\;$ 17.20 & 0.186 $\;$ 0.56\\
 LMC S63 & - $\;$ 467 & - $\;$ 818 & - $\;$ 14.29 & - $\;$ 0.135  \\
 SMP LMC 94 & - $\;$ 289 & - $\;$ 815 & - $\;$ 16.68 & - $\;$ 0.068  \\
  \hline
 \end{tabular}

\end{table*}

\begin{figure*}
\begin{center}
\includegraphics[width=0.22\textwidth]{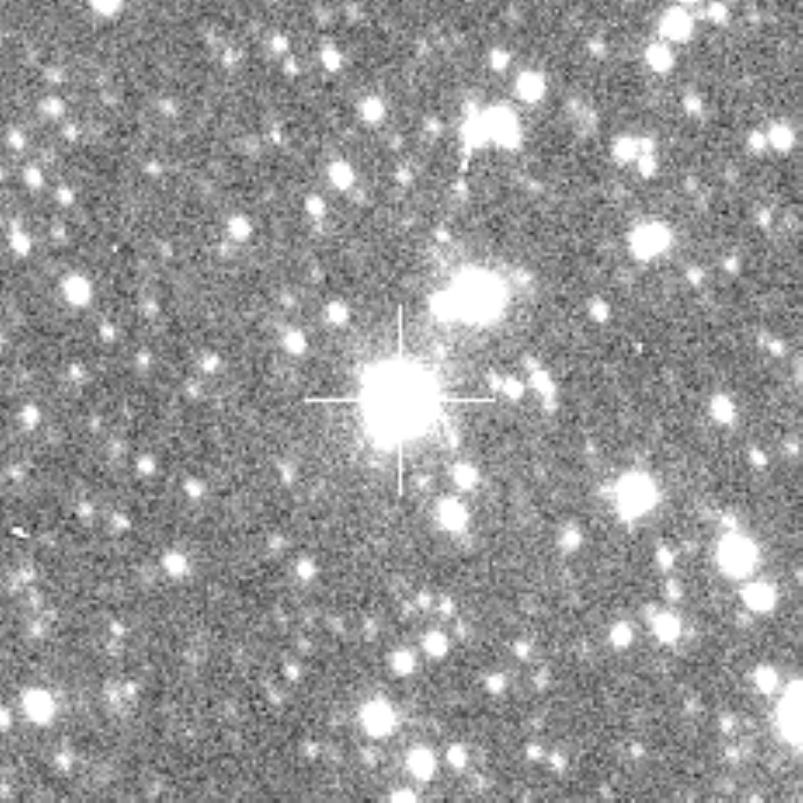}
\includegraphics[width=0.22\textwidth]{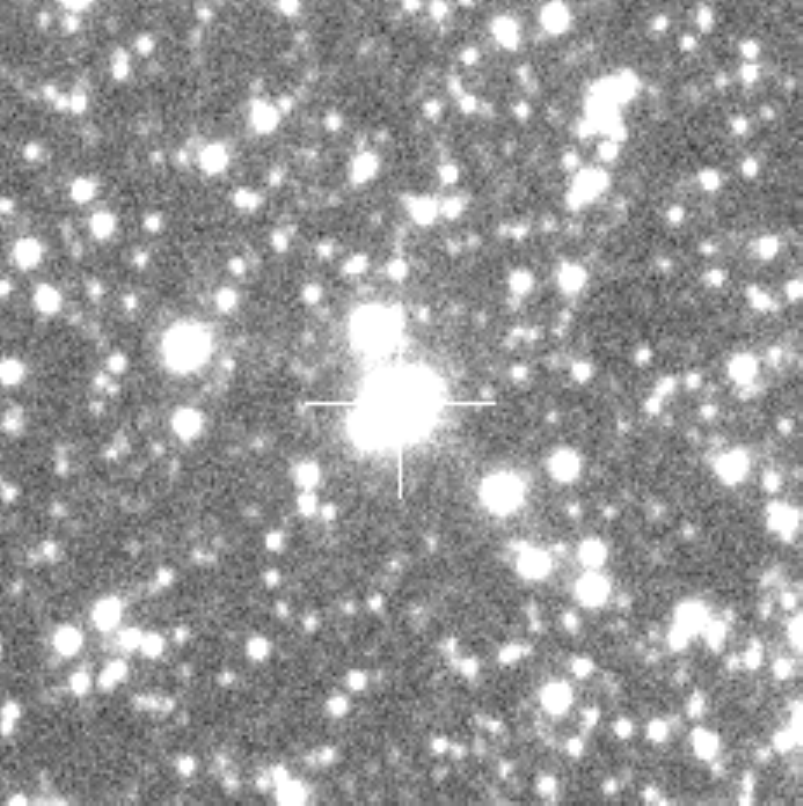}
\includegraphics[width=0.22\textwidth]{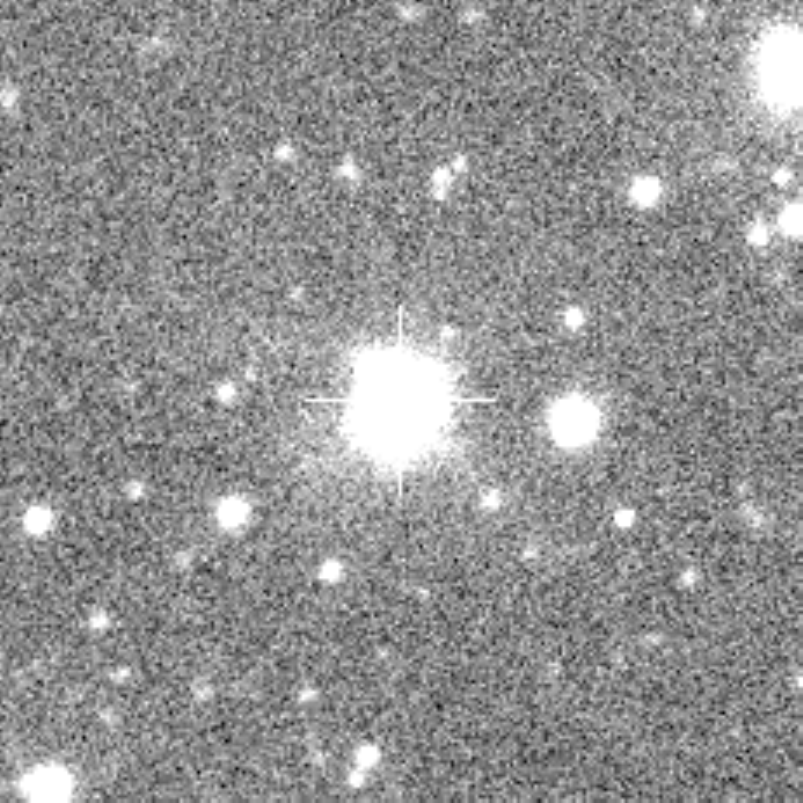}
\includegraphics[width=0.22\textwidth]{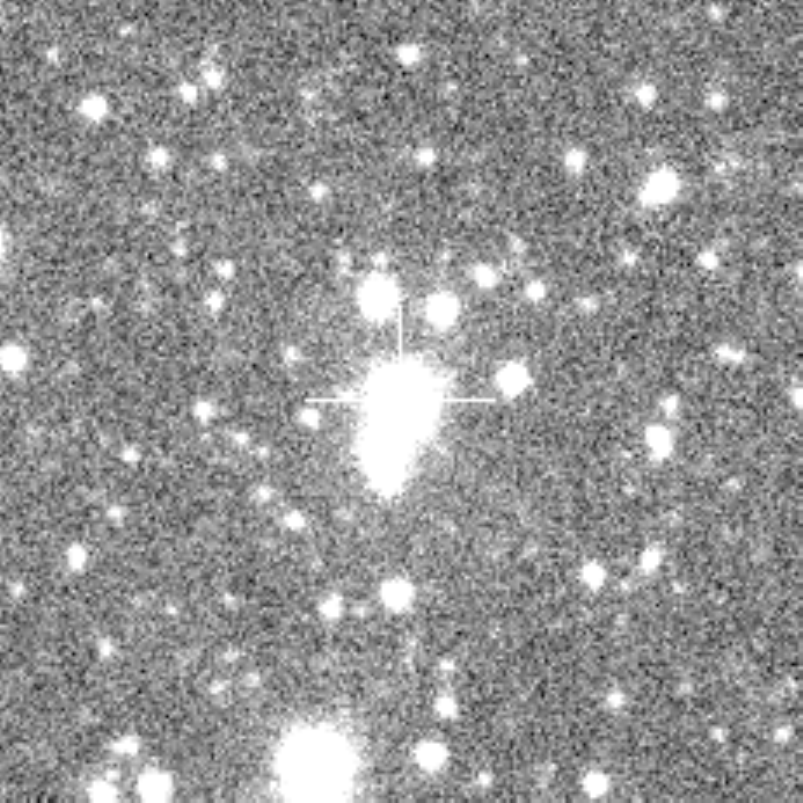}

\includegraphics[width=0.22\textwidth]{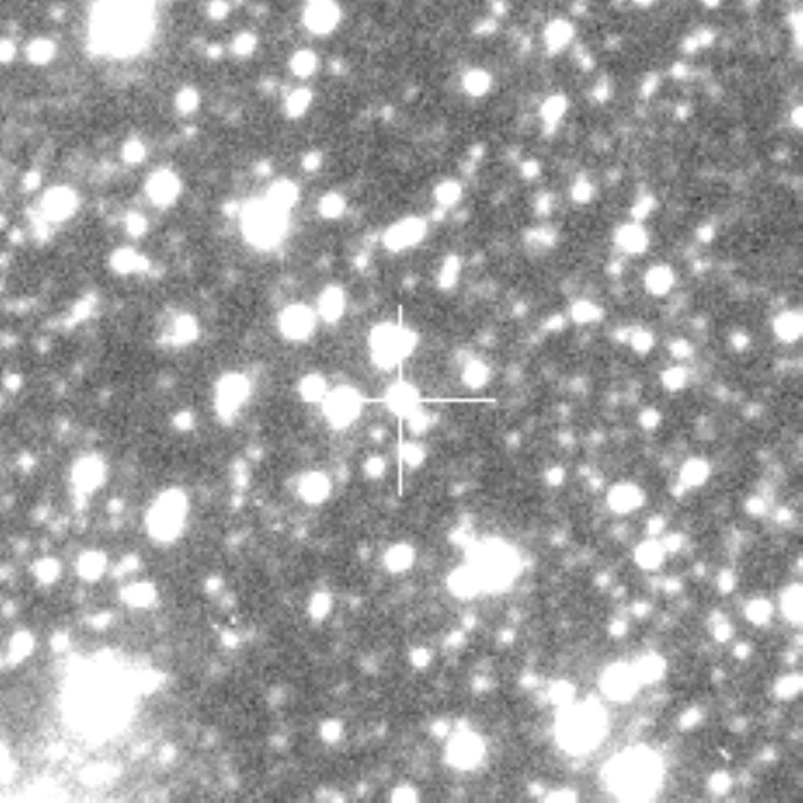}
\includegraphics[width=0.22\textwidth]{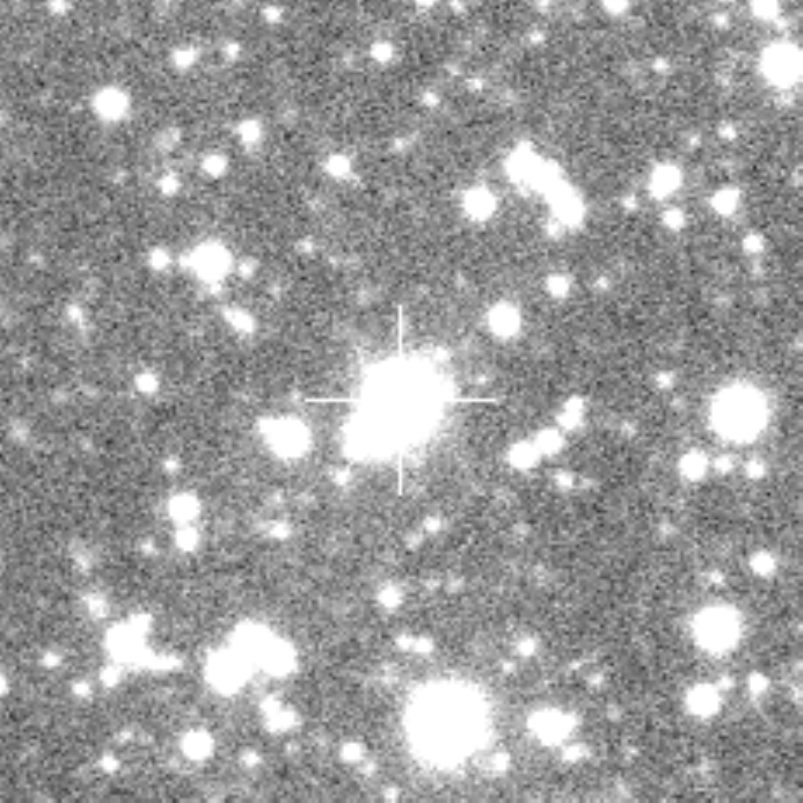}
\includegraphics[width=0.22\textwidth]{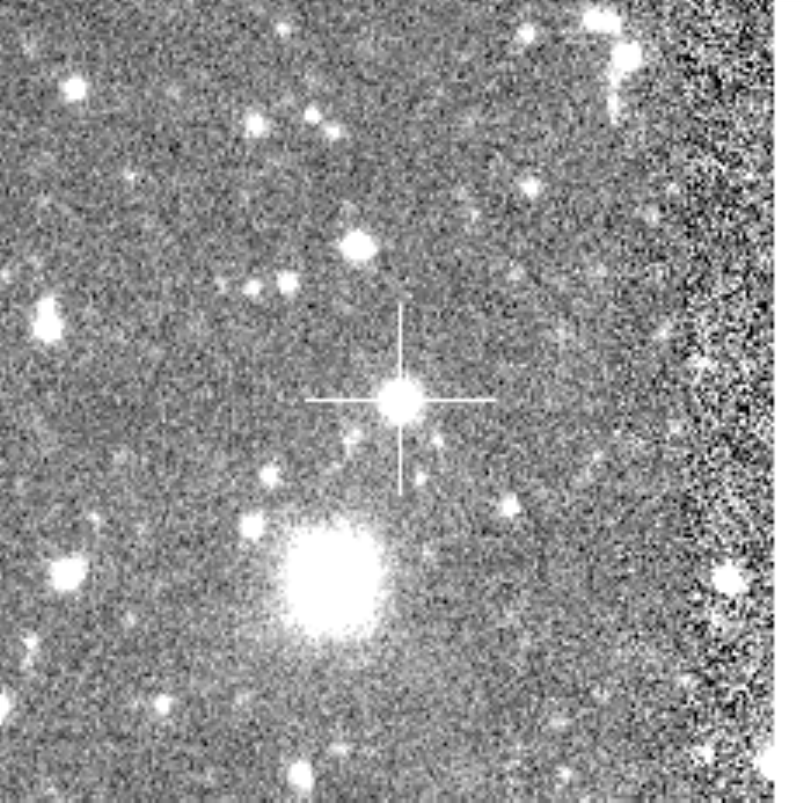}
\end{center}
\caption{OGLE-IV finding charts of LMC SySs. The FoV is 60''x60'' with N up and E left. The target is at the image centre as indicated with a fine crosshair. Top panels, from the left: LMC S147, LMC N19, LMC 1, LMC N67. Bottom panel, from the left: Sanduleak's star, LMC S63, SMP LMC 94.\label{fig:fc}}
\end{figure*}

\subsection{Complementary data}\label{sec:data}
In the MACHO Project Data Archive, we recovered unpublished data of Sanduleak's star. This long-term light curve is complementary to the OGLE data both in wavelength and temporal coverage. The MACHO Project observed Sanduleak's star throughout the 1990's with B and R filters.  (Their filters are non-standard, see e.g., Alcock et al. 1992.)  When combined with the OGLE-III and -IV data, it results into a continuous multi-band monitoring of Sanduleak's star over more than 20 years. In Sec. \ref{sec:sandy} we will further discuss the significance of this unique data-set.

For Sanduleak's star, we have also searched all deep archival plates in the collection at the Harvard College Observatory, spanning a time from 1896 to 1951, with limiting $B$-band magnitudes from 17.0 to 18.0.  Sanduleak's star was not seen on any of these plates, and this will show that the trend over the last 20 years is not a continuation from earlier times.

\subsection{Data analysis}\label{sec:analysis}
SySs are composite astrophysical systems that show marked variability at basically all time scales and wavelengths. It is therefore quite natural that this variability appears composite and intrinsically multi-periodic.  A consequence is that there is no unique or standard recipe to extract the maximum informative content from a symbiotic light curve.

To start, we make a careful visual inspection of the dataset, picking out the most prominent periodicities and characterize the general properties of the light curve.  This allows us to recognize any changes that are not strictly periodic, including short-term random variability, active/quiescent states, and secular trend.

Since we are dealing with unevenly-spaced data, we have paid particular attention to the time distribution of magnitudes in the light curve.  This data windowing depends only on the time sampling (with daily gaps, cloud gaps, lunar gaps, and yearly gaps in data taking) and not upon the signal itself.  The problem is that spurious periods can arise from the gaps present in time-series data. If $P$ is the true period and $P_{\rm gaps}$ is a period of the gaps, spurious periods are then given by:
$$P_{\rm spur}^{-1} = P^{-1} \pm k \, P_{\rm gaps}^{-1} $$ 
with $k$ is some small integer (Lafler \& Kinman, 1965).  

Identifying such periods is of course an important preliminary step needed to flag any aliasing that might mislead the physical analysis.   We have then made use of various time-series analysis techniques mainly based on Fourier decomposition and statistical methods (see, e.g., Templeton 2004 for a general review): in particular, the String-Length Minimization (SLM) method, which was developed by Lafler \& Kinman (1965) and improved by Stetson (1996); the generalized Lomb-Scargle periodogram (GLS - Lomb 1976, Scargle 1982, Zechmeister \& Kumlrster 2009); and the Phase Dispersion Minimization (PDM - Stellingwerf 1978).  These methods require the user to first specify the frequency range $f_{0}$, $f_{N}$ and resolution $\Delta f$ of the period search. Since our dataset is quite heterogeneous in terms of temporal baseline and number of data point per object (Table \ref{tab:oglemasys}), we have always adopted a consistent set of search parameters that depend on the observational data set.  For the low-frequency 
limit we adopt the value of $f_{0} = 0.5 / T_{\rm tot}$ day$^{-1}$, where $T_{\rm tot}$ is the total time span of the observations.  For the high-frequency limit, we adopt $f_N = 0.5$ day$^{-1}$, so as to avoid the periods of $1$ day and its aliases.  For the frequency resolution, we adopt $\Delta f = 0.1 / T_{\rm tot}$ day$^{-1}$.

It has been shown that these different methods are similarly sensitive to features in the power spectrum.  For example, Di Stefano et al. (2012) have shown that for solar-like stars with irregularly sampled datasets, the GLS method is more efficient to detect periods $<$20~d, whereas SLM is slightly more efficient for periods $>$20~d. In our case, there has been a general agreement between the different methods in recognizing the most prominent periodicities in the power spectra, out of which we have decided to keep only the strongest two that could also be positively interpreted in terms of physical and/or geometrical effects of the symbiotic phenomenology. We eventually performed also \textit{a posteriori} visual inspection to cross-check whether the periods recovered via time-series analysis could consistently fit into the overall physical interpretation we were proposing also on the basis of the (e.g., spectroscopic) data from the literature (see, for example, Section \ref{s147}
 for the case of LMC S147).

\section{Results}
Table \ref{tab:masys} presents the names and coordinates of the seven studied SySs, along with the corresponding field of each object in the OGLE-III and -IV surveys. Finding charts for all our stars are displayed in Fig. \ref{fig:fc}.  Table \ref{tab:oglemasys} gives a general overview of the OGLE observations, by summarizing for each object and filter the total number of data points in the light curve, the overall time baseline, the average magnitude, and the maximum amplitude of variability. For the two targets (i.e., LMC N19 and Sanduleak's star) observed during the third phase of the OGLE Survey, our time baseline is more than ten years with a I-band average sampling of about one visit every five days. The remaining five objects have been monitored starting with OGLE-IV, namely, for a bit more than three years at the time of writing. For our target stars, the cadence is generally higher than in OGLE-III, in some cases being as high as one visit every two days.\\

SySs light curves have complex profiles as direct consequence of their composite nature. So, while the $V$-band is more sensitive to the activity of the symbiotic nebula, the $I$-band is mainly dominated by the cool giant.  The nebula and any accretion disk (whenever present) can still have some non-negligible effects, especially during active states.  Gromadzki et al. (2009) has shown that red/IR light curve variations in galactic D-type SySs are usually dominated by the AGB pulsations.  The exception is that circumstellar dust extinction might hide the underlying photosphere, though in this case, the light curve will show slow secular changes due to dust reprocessing and/or nebular evolution. In the case of S-type SySs, the distinctive feature of red light curves are still the pulsations of the cool giant variable and, if the binary is eclipsing or the giant is ellipsoidally distorted, one may even indirectly estimate the orbital period (Miszalski et al. 2013). We expect our $I$-band light curves 
from OGLE 
to be most useful to constrain the nature of the giant pulsator, reveal any ellipsoidal modulation, and detect any dust obscuration.  Our $I$-band data are typically much-less sensitive to reflection effects, which rapidly become undetectable when moving towards the red, even in cases of high-amplitude variations in the U and B bands. This is unfortunate because reflection effects have proved useful in finding orbital period for SySs (see the exemplary case of YY Her - Wiecek et al. 2010).\\

In the following subsections, we present the OGLE light curves for the individual object of the sample and discuss the results of our investigation based on the methodology detailed in Section \ref{sec:analysis}.

\subsection{LMC S147}\label{s147}
LMC S147 was spectroscopically identified and classified as a SyS by Morgan \& Allen (1988).
M\"{u}rset et al. (1996) reported optical/NIR magnitudes and colours for the system, ultraviolet spectroscopy (acquired with {\it IUE} in 1993),  and emission parameters (such as luminosity, radius and  temperature) for the two components.  The companion is an oxygen-rich red giant star of K5-M1 spectral type (Morgan \& Allen 1988; M\"{u}rset et al. 1996).  On the basis of the NIR colours, the source is usually classified as a S-type SyS. \\

Mikolajewska (2004) reported on a modulation of $\sim$1000d visible in the MACHO blue channel data, and the LMC S147 power spectrum shown in her Fig. 1 clearly reveals three marked peaks at $\sim$1500, $\sim$635, and $\sim$465 days. The author however lacked the observations needed to confirm the orbital origin of these modulations.\\

\begin{figure*}
\begin{center}
\includegraphics[width=0.90\textwidth]{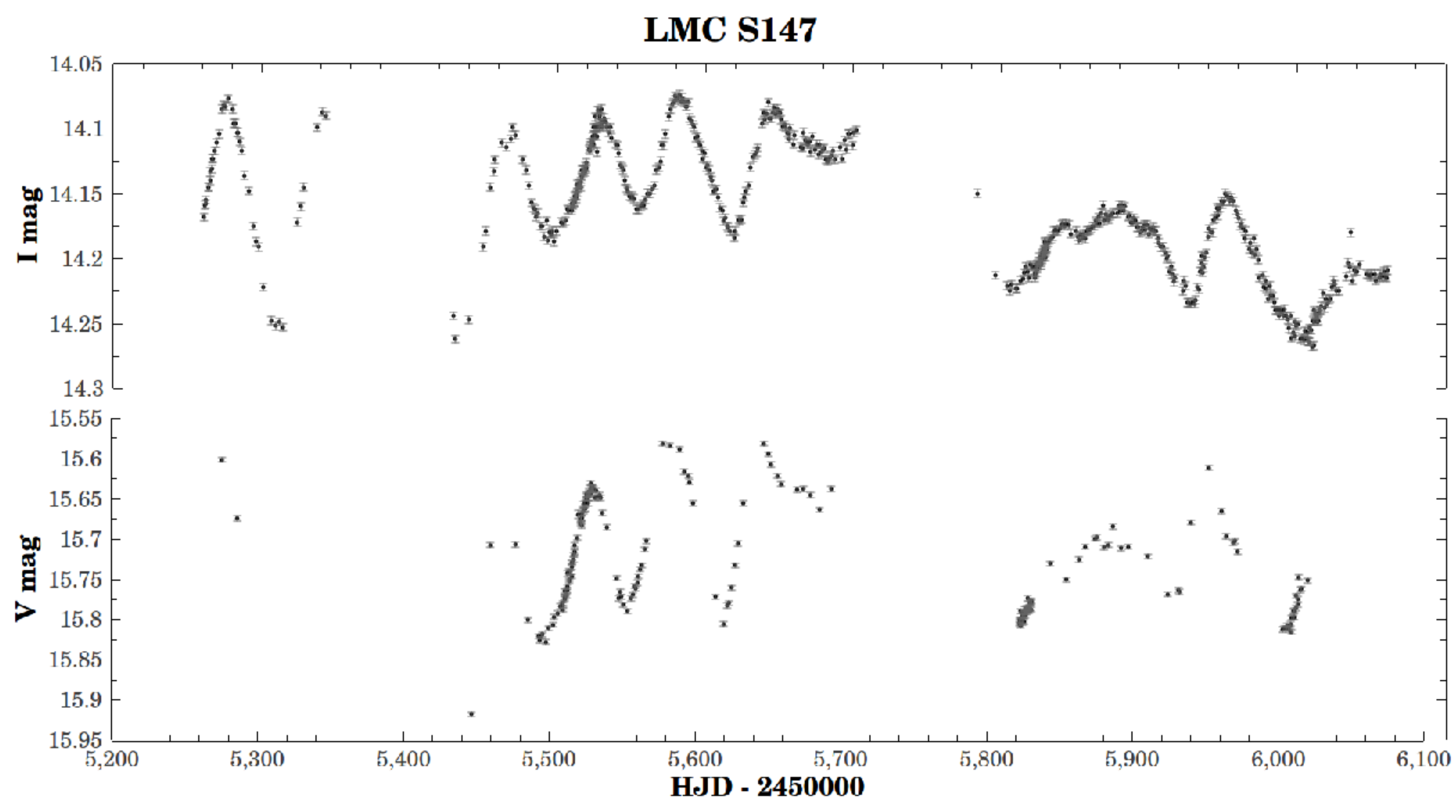}
\end{center}
\caption{OGLE-IV light curves of LMC S147.\label{fig:lmcs147_lc}}
\end{figure*}

The OGLE-IV light curve of LMC S147 is shown in Fig. \ref{fig:lmcs147_lc}. The I band shows a relatively small amplitude variation marked by a clear periodicity, that seems to become more disturbed after HJD~2455800. The light curve analysis identifies this periodicity as the primary period at $P_1=63\pm3$ days. A second periodicity appears at $P_2=248\pm10$ days. We believe this second peak is real because the corresponding \textit{False Alarm Probability} (FAP, Scargle 1982) is $<$0.01. While a reasonable interpretation could relate the primary period to the pulsation of the red giant, it is more critical to associate $P_2$ to any orbital modulation, especially when remembering the periodicity found in the MACHO blue channel data, that are supposedly more sensitive to orbital effects.\\

\begin{figure*}
\begin{center}
\includegraphics[width=0.9\textwidth]{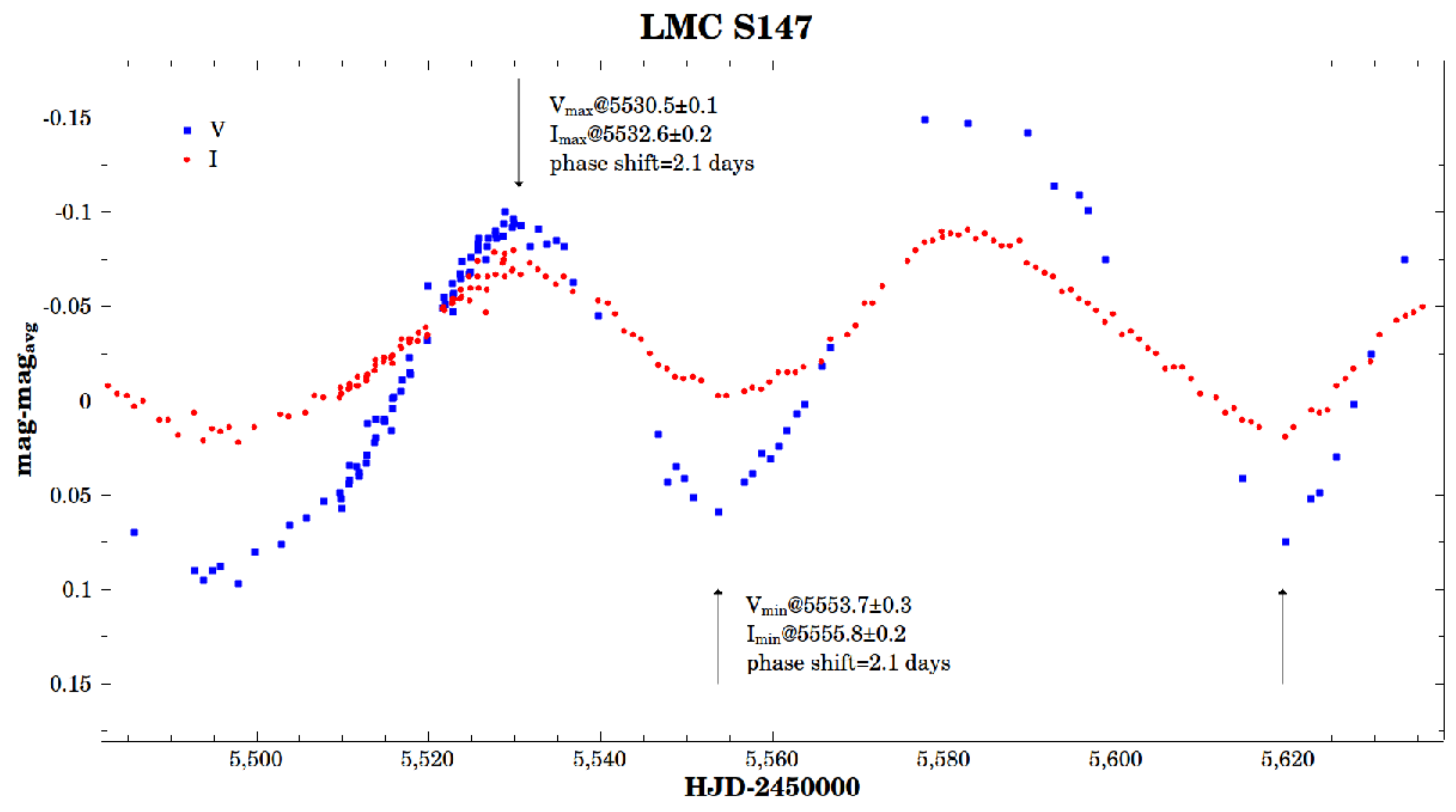}
\end{center}
\caption{Zoom onto LMC S147 between HJD2455450 and -5650. Blue squares and red circles represent average-subtracted V and I magnitudes, respectively. The arrows mark the position of the two minima and one maximum used to calculate the phase shift and the (V-I) colour and thus link the 63-day period to the red giant pulsations. \label{fig:s147_zoom}}
\end{figure*}

To examine whether $P_1$ is related to stellar pulsation phenomena, we concentrate our analysis on that portion of the light curve where this periodicity emerges more clearly (between HJD~2455450 and -5650) and for which we can also exploit the OGLE $V$-band data that thus provide us with essential colour information. A zoom on this time interval is shown in Fig. \ref{fig:s147_zoom}, where we have plot the average-subtracted I and V magnitudes as red circles and blue squares, respectively.\\

After having calculated with the modified Kwee \& van Woerden's Algorithm (Kwee \& van Woerden, 1956) the timing of the well-sampled maxima and minima (whose positions is marked with arrows in Fig. \ref{fig:s147_zoom}), we have started noticing that at V minimum (maximum) the V-I colour is redder (bluer); moreover, the phase shift between the two bands at both the maxima and minima seems to be constant within the algorithm fitting errors, as equally constant is the colour term (V-I=1.63 mag) at the two consecutive minima HJD~24555553 and -5619. All this evidence together indicates the physical origin of the 63-day period is pulsations of the cool component.  A simple linear ephemeris for the time of minima is
$$HJD(min) = 2455553(\pm1 ) + E \times 63(\pm3)$$, where $E$ is an integer to identify the epoch.\\

With $P_1$ linked to the red giant, we are in the position to further constrain its evolutionary status. The average I magnitude of the system is by 0.4 mag brighter than the tip of the RGB in LMC (I=14.56 mag - Udalski 2000), thus implying that the star is currently in the AGB phase. By combining the 2MASS magnitudes as reported in Phillips (2007) with both the OGLE photometry and our period analysis, we can place LMC S147 in the period-luminosity diagrams (see Fig. 1 of Soszy\'nski et al. 2009).  The cool component lies close to the border between OGLE Small Amplitude Red Giants (hereafter OSARG) and oxygen-rich semi-regular variables (hereafter SRV). In general terms, the overall light curve shape is reminiscent of an OSARG (Soszy\'nski et al. 2004), even if the $I$-amplitude is slightly larger (by only 0.06 mag, though) than usually assumed for these pulsating stars ($\Delta I<$0.13 mag).  However, there might still be a small but not negligible contribution of the symbiotic nebula to the $I$-band, so we 
favour the interpretation of the red giant in LMC S147 to be an oxygen-rich OSARG. The physical mechanism that causes the pulsations in OSARGs is still not clear, with alternative theoretical scenarios described by Dziembowski et al. (2001) and Christensen-Dalsgaard et al. (2001).

\subsection{LMC N19}
LMC N19 was first identified by Morgan (1996) during a visual scanning of an objective-prism plate taken with the UK 1.2m Schmidt Telescope (UKST) at Siding Spring, Australia. The spectrum revealed a typical SyS with a M4 giant and strong, high-excitation emission lines.  No forbidden ones appeared, as it is the case for S-type SySs.\\

Curiously, no photometric variation could be detected, even with additional UKST B,V,R and I plates (Morgan 1996).  However, Morgan says that he found LMC N19 brighter by $\sim$2 mag in U band plates from the 1980's with respect to a similar plate taken in 1977.  This failure in detecting any longer wavelength variability is quite surprising, because in more than 10 years of OGLE monitoring, LMC N19 has exhibited an overall amplitude of almost 0.8 and 0.5 magnitude in V and I, respectively (Table \ref{tab:oglemasys}).\\

Mikolajewska (2004) analysed the MACHO light curve of LMC N19 and reported two stronger periodicities at $\sim$1000 days and $\sim$80 days.  She interpreted these as likely indications of the  binary orbital period and radial pulsation of the M4 component, respectively.\\

\begin{figure*}
\begin{center}
\includegraphics[width=0.90\textwidth]{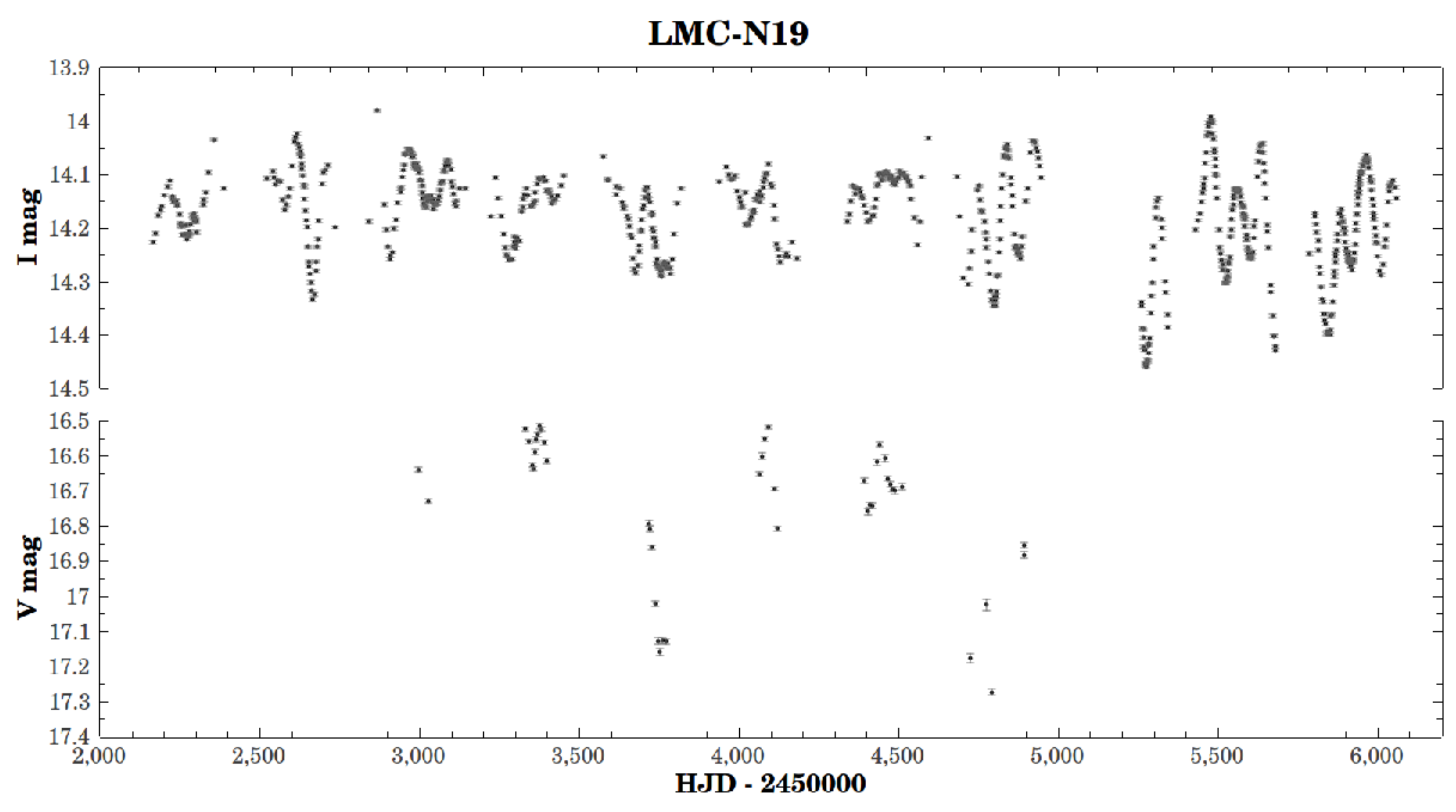}
\end{center}
\caption{OGLE light curves of LMC N19.\label{fig:lmcn19_lc}}
\end{figure*}

\begin{figure*}
\begin{center}
\includegraphics[width=0.90\textwidth]{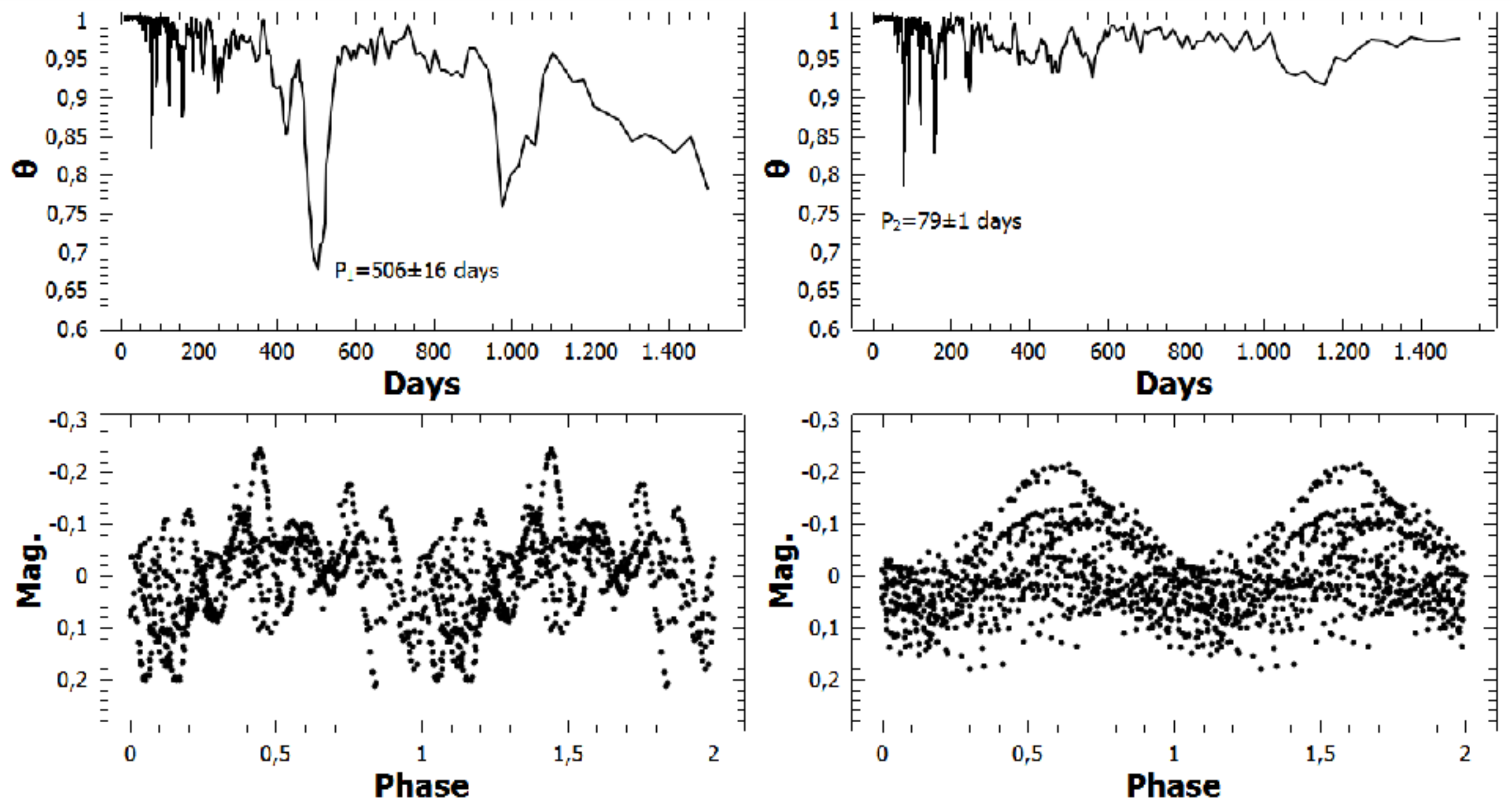}
\end{center}
\caption{Results of PDM analysis for LMC N19. Top panels: periodograms showing the primary period P$_1$=506$\pm$16 days (left panel) and the secondary period P$_2$=79$\pm$1 days, after prewhitening with P$_1$ (right panel). Bottom panels: the light curve phased with P$_1$ (left panel) and P$_2$ (right panel). \label{fig:lmcn19_periodogram}}
\end{figure*}

In Fig. \ref{fig:lmcn19_lc}, we present the V and I light curves obtained during the third and fourth phase of the OGLE Project. The character of the well-sampled $I$-band light curve is clearly multi-periodic, and the period analysis reveals a first periodicity at P$_1$=506$\pm$16 days, and a second one at P$_2$=79$\pm$1 days (Fig. \ref{fig:lmcn19_periodogram}). The identification of P$_2$ with the stellar pulsation of the giant component is quite natural.  This also agrees fairly well with the period deduced from the MACHO red channel data.  Our P$_1$ is about half the period suggested by Mikolajewska as deduced (presumably) from the MACHO blue channel data.  We note that the ratio 1:2 between P$_1$ and the $\sim$1000 days period, if physically real, suggests an ellipsoidal distortion of the giant component, and hence the orbital period would be $\sim$1000 days.  We cannot rule out, nor surely confirm, this possibility on the basis of our OGLE I band data. Within this idea, the $\sim$2 mag $U$-band 
variation reported by Morgan (1996) might be explained as a poorly-covered active states of the system.\\

Taking into account the overall $I$-band amplitude, the pulsation period P$_2$=79$\pm$1 days, and the system IR colours (c.f. Phillips 2007), we can classify the giant component in LMC N19 as an oxygen-rich SRV probably pulsating in its first overtone (Soszy\'nski et al. 2007). An ephemerides for the giant pulsations is
$$HJD(min) = 2456012(\pm1 ) + E \times 79(\pm1)$$,
where E is an integer.

\subsection{LMC 1}
SyS LMC 1 was also discovered by Morgan (1992) during visual scans of objective prism plates taken with UKST. The object appeared as a fairly normal carbon star with a rich emission line spectrum dominated by the Balmer series and by a large number of forbidden lines in which the particularly strong [OIII], [FeV] and [FeVI] emission lines were reminiscent of the galactic symbiotic nova RR Tel.  A few years later M\"{u}rset et al. (1996), as based on {\it IUE} spectra, estimated a temperature of LMC 1 hot component as high as 125,000 K.  On the basis of the NIR colours, they classified LMC 1 as a D-type SyS.\\

\begin{figure*}
\begin{center}
\includegraphics[width=0.90\textwidth]{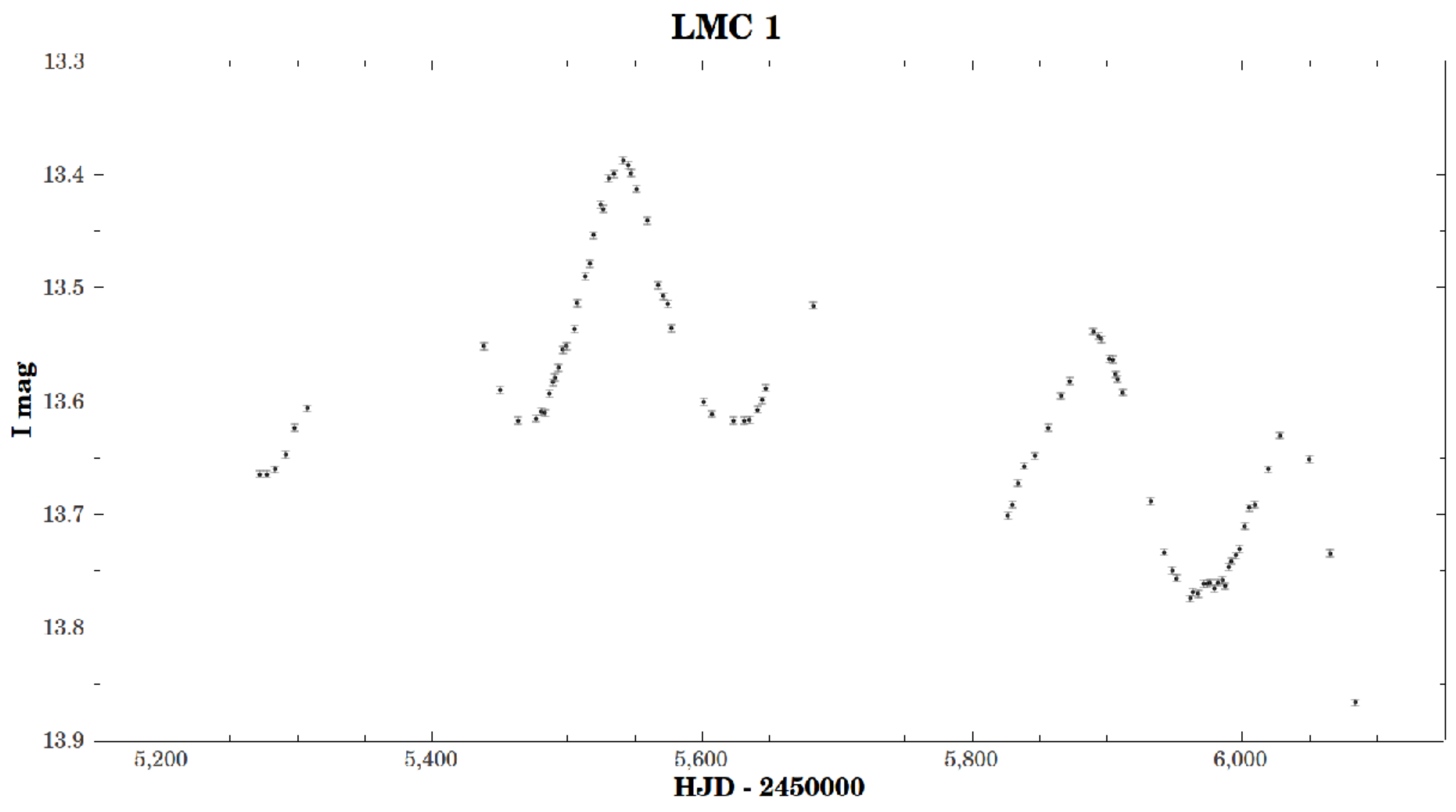}
\end{center}
\caption{OGLE-IV light curve of LMC 1.\label{fig:lmc1_lc}}
\end{figure*}

The $I$-band OGLE-IV light curve of LMC 1, shown in Fig. \ref{fig:lmc1_lc}, hides the strong symbiotic character of this binary system under the regular periodicity of its cool component. Despite the seasonal gaps visible in the data distribution, an overall amplitude of almost half a magnitude and a marked periodicity at $\sim$170 days are clearly recognizable with a visual inspection.\\

\begin{figure*}
\begin{center}
\includegraphics[width=0.9\textwidth]{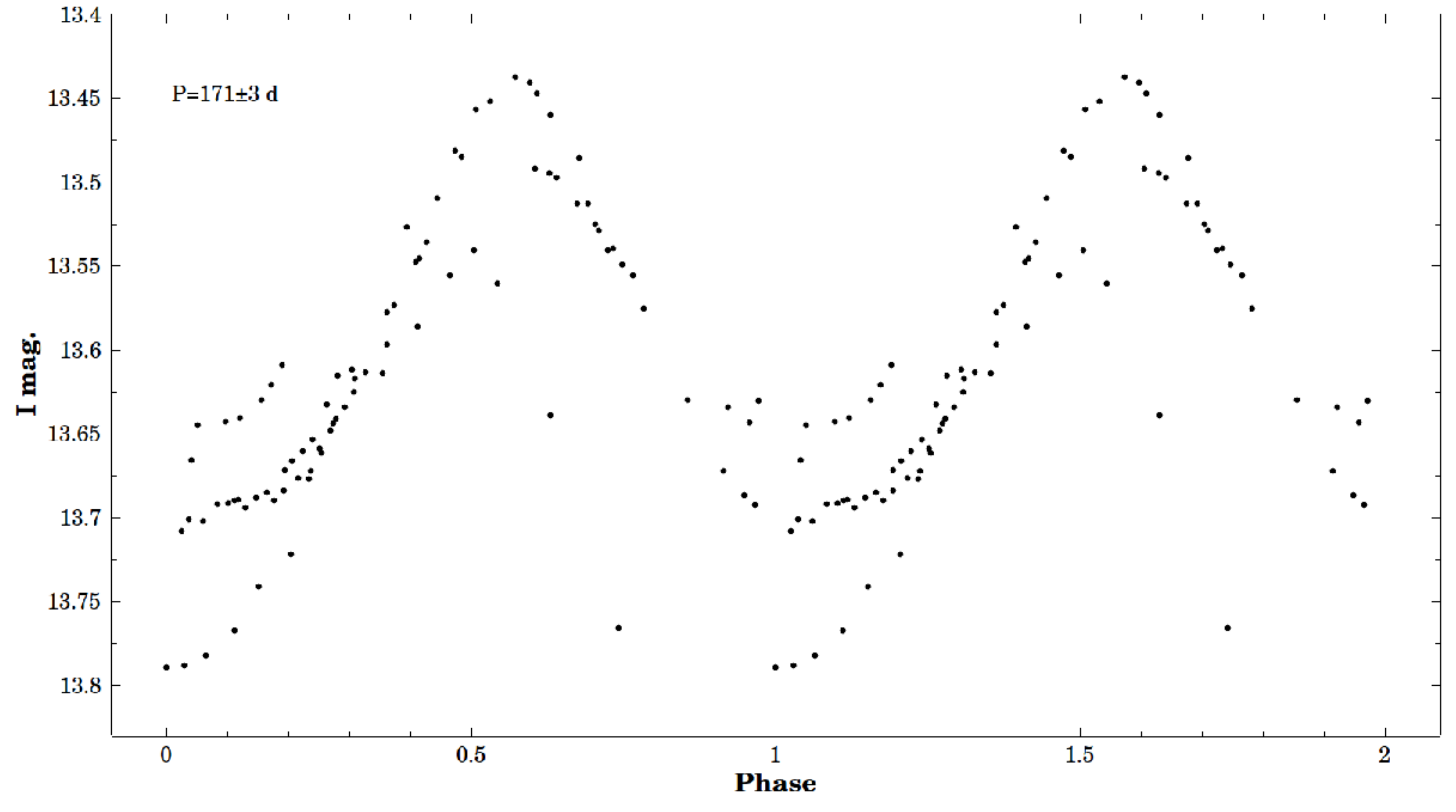}
\end{center}
\caption{The light curve of LMC 1 folded with the primary period P$_1$=171$\pm$3 days\label{fig:lmc1_phase}.}
\end{figure*}

The GLS analysis reveals, along with the primary period P$_1$=171$\pm$3 days, also a second significant period with P$_2$=98$\pm$2 days.  This is the likely origin of the scatter in the phase plot of Fig. \ref{fig:lmc1_phase} folded with P$_1$. Our linear ephemeris for LMC 1 is
$$HJD(min) = 2455962(\pm1 ) + E \times 171(\pm3)$$.
When we use the primary period (from our analysis) and 2MASS JHK colours (Phillips 2007) to locate LMC 1 in the diagnostic period-luminosity diagram (Soszy\'nski et al. 2007), we are able to recover the carbon-rich nature of its giant.  We also conclude that it is probably a SRV pulsating in its first overtone.\\

A careful visual inspection of light curve profile seems further to suggest that no ellipsoidal variation is present, thus implying that the carbon giant in LMC 1 is not tidally distorted. A direct physical interpretation is that the mass accretion in LMC 1 is from a stellar wind.\\

We finally notice that the seemingly modulated difference in the brightness level of the three recorded maxima could be a long secondary period (LSP), feature observed in at least 30\% of known red giants (Soszy\'nski et al. 2004). No acceptable explanation for LSPs in AGB stars has been provided (Nicholls 2011). In the case of LMC 1, more data are clearly required to tell whether this LSP may stem from dust obscuration events or is related with orbital motion. In the latter case, the resultant lower limit on the orbital period would be P$_{orb}>$1500 days, and this would locate LMC 1 in the long-period tail of galactic SySs (Mikolajewska 2003).

\subsection{LMC N67}
The symbiotic classification of LMC N67 was due to Cowley
\& Hartwick (1989), who also identified the
secondary star as a carbon star. However, Morgan (1992) suggested that this object
may be carbon poor. The spectral type of the secondary was 
given as C3.2 by M\"{u}rset et al. (1996), who also give 
optical/NIR magnitudes, colours and other information on the emission 
properties of the system. The object was reported by 
Belczy{\'n}ski et al. (2000) as an S-type SyS with 
$V$ = 15.9 and $V-K$ = 4.5 mag. \\
An {\it IUE} ultraviolet spectrum of the source(Vogel \& Morgan 1994) shows several emission lines, 
yielding the temperature and luminosity of the hot
component, as well as abundance ratios for the system, which appear similar 
to those of Galactic SySs.\\

\begin{figure*}
\begin{center}
\includegraphics[width=0.90\textwidth]{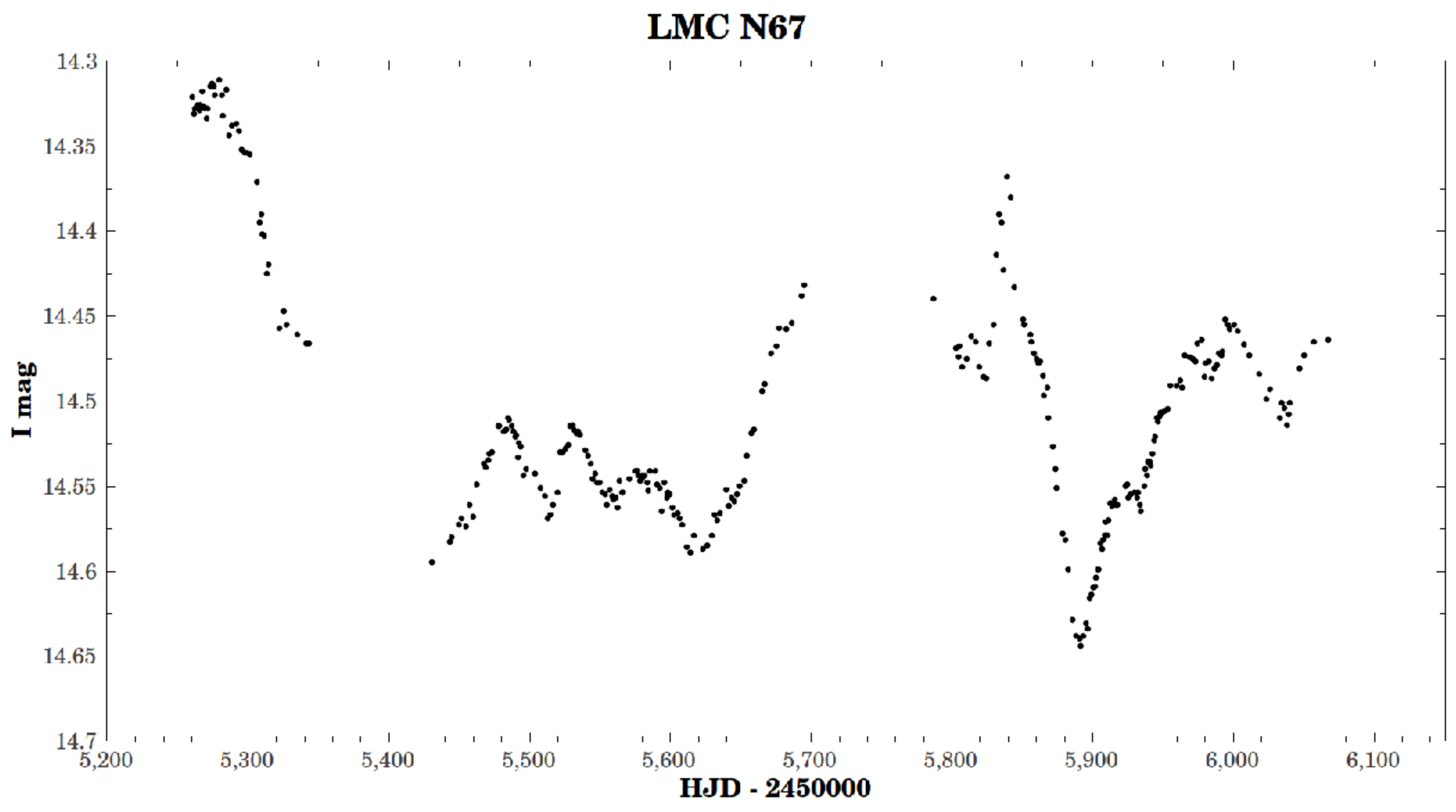}
\end{center}
\caption{OGLE-IV light curve of LMC N67.\label{fig:lmcn67_lc}}
\end{figure*}

The OGLE-IV light curve (Fig. \ref{fig:lmcn67_lc}) presents very interesting features.  After a monotonic dimming by more than 0.2 mag, semi-regular pulsations (with P$\approx$45 days) appear between HJD~2455450 and -5650. Then, a moderately slow rise to a high state, but then our observations were cut off by the seasonal gap.  After this gap, the system enters what we are tempted to identify as an eclipse.  During the egress from the supposed eclipse, it then seems that the semi-regular pulsations gradually reappear, after having ceased in the ingress phase. If this interpretation were correct, by associating the semi-regular variations between HJD~2455450 and -5650 to the pulsations of the symbiotic giant component, this would imply that it is the giant itself that becomes eclipsed.  Unfortunately, the absence of any colour information does not allow us to say whether we are dealing with a dust obscuration episode (quite unlikely, though, the system being supposedly an S-type) or if the photons from the 
giant star are actually absorbed and reprocessed by the symbiotic nebula.  We do not recognize further trends that might correlate with orbital modulations.  Further data (including longer monitoring and colour information) are required to confirm that we are actually dealing with an eclipse.\\

With the IR colours of the system as given in Phillips (2007), and our apparent pulsational period of $\approx$45 days, LMC N67 cool component is probably an OSARG. Our ephemerides is 
$$HJD(min) = 2455513(\pm1) + E \times 45(\pm5)$$
for the minima of the pulsations.\\

\subsection{Sanduleak's star\label{sec:sandy}}
In 1977, N. Sanduleak reported on the discovery of a peculiar emission-line object in the direction of LMC (Sanduleak 1977).  This star is known as `Sanduleak's Star'.  The initially suspected spectral variability, then confirmed by later studies (e.g., Morgan et al. 1992), already suggested that the source had to be ``some type of eruptive variable star rather than a planetary nebula'' (Sanduleak 1977).  The nature of its central star is still controversial. It is usually classified as a SyS, but this classification relies primarily on the detection of the Raman scattered emission features in its spectrum (Munari \& Zwitter 2002).  However, no direct evidence exists in the optical to confirm the presence of the giant component - an essential ingredient of a symbiotic recipe.  Also, its near-infrared colours appear rather peculiar amongst SySs (M\"{u}rset et al. 1996).  In addition, Sanduleak's star has been reported to possess an unusual chemical composition that includes strongly enhanced N/C and N/O 
abundance 
ratios very similar to those observed in the outflows of SN~1987A and $\eta$ Car (Kafatos et al.~1983; Michalitsianos et al.~1989). \\

Virtually forgotten for some time after the discovery, Sanduleak's star has come back to prominence when Angeloni et al. (2011) discovered that it is powering the largest known stellar jet (14 parsecs long) and the first stellar jet outside of the Milky Way to be clearly resolved.  Given this extraordinary jet, along with the precise knowledge of the distance to the LMC (that makes it possible to derive absolute estimates of the physical quantities of the system), we believe that Sanduleak's star can turn into a critical test-bed for theoretical modelling of astrophysical jets. The upcoming HST Cycle 20 observations (GO-12868, PI: Angeloni) are expected to largely improve our understanding of this extreme stellar structure. \\

Galactic SySs with jets (e.g., CH Cyg, see Karovska et al. 2010) are best detected in the X-ray region, so we would expect that the incredibly long jet from the Sanduleak's star to also be X-ray luminous.  Nevertheless, Sanduleak's star has never been detected in X-rays. Bickert et al. (1996), from data acquired in September 1990 and February 1991, report {\it ROSAT}/PSPC 3-$\sigma$ upper limits of (1.0--3.2)$\times$10$^{-3}$ cts s$^{-1}$ in the 0.1--2.4 keV band. This corresponds to (5.6--18)$\times$10$^{-15}$ erg cm$^{-2}$ s$^{-1}$. A comparable upper limit (2.1$\times$10$^{-3}$ cts s$^{-1}$ or 1.2$\times$10$^{-14}$ erg cm$^{-2}$ s$^{-1}$) in September 1993 was obtained by M\"{u}rset et al. (1997), again with {\it ROSAT}/PSPC data.  We have analysed X-ray data from two other spacecrafts:  (1) We used unpublished 0.3--10 keV data from the XRT detector (Burrows et al. 2004) onboard the {\it Swift} satellite (Gehrels et al. 2004), with this being available in the public archive. The position of 
Sanduleak's star 
was indeed serendipitously observed for 2740 s between 28 and 29 December 2008 during a pointing of the nearby source CAL 87. The XRT data reduction was performed using the XRTDAS standard data pipeline package ({\tt xrtpipeline} v. 0.12.6), to produce screened event files. All data were extracted in the photon counting (PC) mode (Hill et al. 2004), adopting the standard grade filtering (0--12 for PC) according to the XRT nomenclature.  Zero X-ray counts were detected in the 0.3--10 keV band within 20$''$ of the position of Sanduleak's star.  Using the Bayesian approach of Kraft et al. (1991), we determined a 99\% count-rate upper limit of $\sim$2$\times$10$^{-3}$ cts s$^{-1}$, corresponding to a 0.3--10 keV flux of $\approx$7$\times$10$^{-14}$ erg cm$^{-2}$ s$^{-1}$.  (2) The source was also not detected in any of the {\it XMM--Newton} observations on this field in the 0.2--12 keV band down to 3-$\sigma$ 
flux upper limits\footnote{available at {\tt http://xmm.esac.esa.int/external/xmm\_products/\\/slew\_survey/upper\_limit/uls.shtml}} of 1.7$\times$10$^{-12}$ erg cm$^{-2}$ s$^{-1}$ and 3.3$\times$10$^{-12}$ erg cm$^{-2}$ s$^{-1}$ in the {\it XMM--Newton} Slew Survey (Saxton et al. 
2008; exposure times of 10 and 5 s acquired on November 8, 2001 and on November 29, 2006, respectively) and of 8.3$\times$10$^{-15}$ 
erg cm$^{-2}$ s$^{-1}$ again during a serendipitous pointed observation of source CAL 87 executed on April 18, 2003 and which
lasted about 22600 s (actual on-source time).\\

\begin{figure*}
\begin{center}
\includegraphics[width=0.90\textwidth]{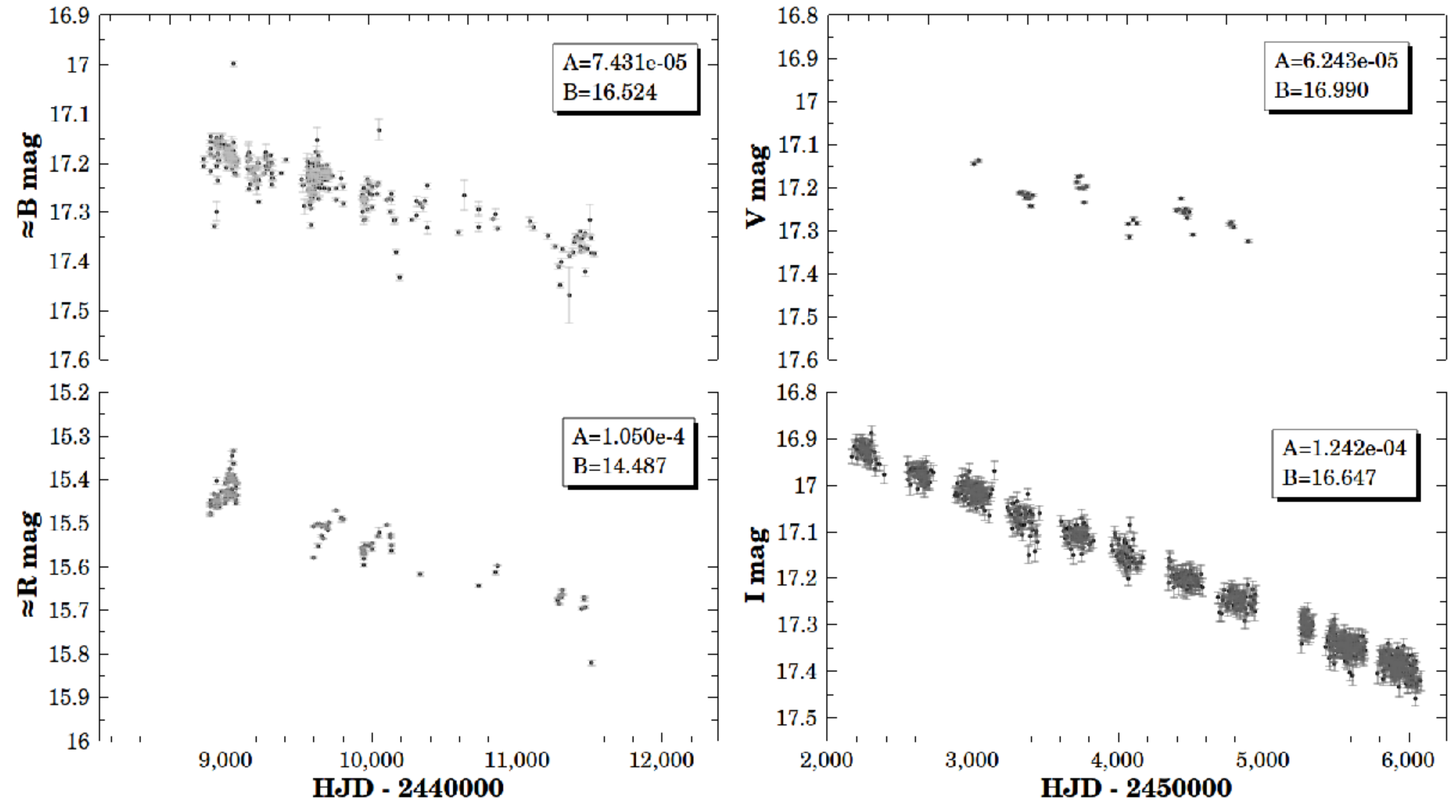}
\end{center}
\caption{MACHO (left) and OGLE (right) light curves of Sanduleak's star. A and B are the coefficients of a weighted linear fit of the data.  The relative slopes among the different filters rule out the possibility that at the origin of the constant-rate fading of Sanduleak's star can be due to a dust obscuration event (see text for details).  An extrapolation of this trend to 1951 suggests a $B$-band magnitude of 16.0, whereas Harvard plates from 1896 to 1951 show upper limits on the brightness from 17.0 to 18.2 mag.  Therefore, the steady linear 20-year trend can only have started in prior decades with the Sanduleak star rising from a much fainter state.
\label{fig:sandy_lc}}
\end{figure*}

By combining MACHO and OGLE data, we have been able to assemble for this star a unique light curve which extends back to the last 21 years over four pass bands, from B to I. For at least two decades, Sanduleak's star has been monotonically fading at an apparently constant rate of $\sim$0.03 mag/year in all bands.  The temporal analysis performed after the usual detrend has not highlighted any significant periodicity in any band.  The constant-rate fading is intriguing.  A dust obscuration event as explanation of the observed fading can be ruled out by the relative slopes of the linear fit across the different wavelengths, because absorption by dust grains would have produced a steeper slope toward shorter wavelengths, but this is not see (Fig. \ref{fig:sandy_lc}). We note that linearly decreasing trends have already been observed in symbiotic novae that were recovering from powerful outburst events (e.g., RR Tel, RX Pup - Gromadzki et al. 2009).  These similar events have 
total radiation outputs of typically $L_{bol}\sim10^{47}$ erg, with the fading due to the decline of both the hot stellar components and nebular radiation (M\"{u}rset \& Nussbaumer 1994).\\

If the observed steady fading at the rate in the $B$-band (0.027 mag/year
from 1991 to 2012) is simply extrapolated into the past,
then we would expect Sanduleak's star to have been
B=16.0 in the year 1950 and B=14.6 in 1900.  We can test
this extrapolation with the collection of archival photographic
plates at the Harvard College Observatory.  We have visually
inspected just over a thousand of the deepest plates showing
the LMC, distributed over the years 1896-1989.  Sanduleak's
star was not seen on any of these plates.  The limiting
magnitudes of the plates were estimated to a typical accuracy
of 0.15 mag, with a sequence of comparison stars based on
the Large Magellanic Cloud Stellar Catalog by Zaritsky et al.
(2004).  The Harvard plates have a zero colour term with respect
to the Johnson B magnitude system, and the use of modern
comparison star magnitudes means that our limits are
exactly in the Johnson B-magnitude system.  In Table \ref{tab:dash}, we
list the plate ID, the year, and the limiting B magnitude for the
deepest plates.  In addition, the Damon plate series has fifty
plates from 1970-1989 that all place limits of B$>$14.8 for
Sanduleak's star.  The complete absence of the star from
the Harvard plates from 1896 to 1989 suggests that it did not
have any major eruptions or flaring events over that century.
(Any such eruptive event must be different from the main
ejection event at the origin of the giant collimated jet, whose
kinematic age has been estimated as  $\sim10^4$ years; see
Angeloni et al., 2011.)  With B=17.15 in 1991 (see Fig. \ref{fig:sandy_lc}),
all of the limits in Table \ref{tab:dash} show that the object was actually
fainter in the prior century, in stark contrast to the last 21
years of steady fading.  In all cases, the current long-lived steady-fading must have started
sometime between 1951 and 1991, and it was preceded
by some sort of a brightening trend or switch to a high state.\\

Finally, we want to point out intriguing similarities between the Sanduleak's star and some explosive stars.  Szczygiel et al. (2012) have recently reported on their discovery that the candidate progenitor of SN2011dh in M51 was fading with a monotonic trend of 0.039$\pm$0.006 mag/year in the time just preceding the SN explosion.  This trend is intriguingly close to Sanduleak's star steady decline.  In addition, the jet in Sanduleak's star has kinematic and chemical similarities with the S condensation of $\eta$ Carinae and with the SN1987A remnant (Michalitsianos et al. 1989).\\

\begin{table}
\centering
 \caption{Magnitude limits for Sanduleak's star from 1896 to 1951.\label{tab:dash}}
 \label{period}
 \begin{tabular}{@{}cccccccccccccccc}
  \hline
   Plate ID & Year & $B$ (mag) \\
  \hline
A2172     &      1896 &   $>$17.8\\
A2202     &      1897  &  $>$17.2\\
A12326    &      1923  &  $>$17.5\\
A14366    &      1929  & $>$18.0\\
A14302    &      1929  & $>$18.0\\
A18022    &      1935  &  $>$17.5\\
A18171    &      1936  &  $>$18.3\\
MF22485   &	 1936  &  $>$17.5\\
A21501    &      1939  & $>$17.9\\
A22980    &      1940  & $>$18.0\\
A22380    &      1940  & $>$18.0\\
A24625    &      1945  &  $>$18.2\\
A26593    &      1948  &  $>$17.7\\
A26998    &      1949  & $>$17.3\\
MF39792   &      1951  & $>$17.0\\
  \hline
 \end{tabular}
\end{table}

\subsection{LMC S63}\label{sec:lmcs63}
This system was first identified as a SyS by Allen (1984). The 
giant companion was spectroscopically classified as a Carbon star (Allen 
1980). M\"{u}rset et al. (1996) refined this as type C2.1J, and also 
gave optical and IR magnitudes.  Belczy{\'n}ski et al. (2000) classified 
it as an S-type system, and measured $V$ = 15.2 mag and 
$V-K$ = 3.9 mag. Gaposhkin (1970)
reported that the light curve of the system bears similarities with that 
of R CrB, although doubts on this are cast by Lawson et al. (1990).
More recently, Hedrick \& Sokoloski (2004) detected variability in B- and V-band over time scales of tens of days from LMC S63, while Fraser et al.
(2008) found an optical periodicity of $\approx$1260 days and amplitude
$\sim$0.3 mag from MACHO data.\\

Ultraviolet spectroscopy with {\it IUE} was acquired on March 1982
(Kafatos et al. 1983), with the spectrum showing numerous emission lines, from which they derived
physical parameters of the hot component along with those of the nebular 
gas (see also
M\"{u}rset et al. 1991). Further ultraviolet spectra were acquired in 1994 
with {\it IUE} and {\it HST} (M\"{u}rset et al. 1996; Vogel \& Nussbaumer 
1995), which allowed a better determination of the main physical parameters 
of the hot component.\\

\begin{figure*}
\begin{center}
\includegraphics[width=0.90\textwidth]{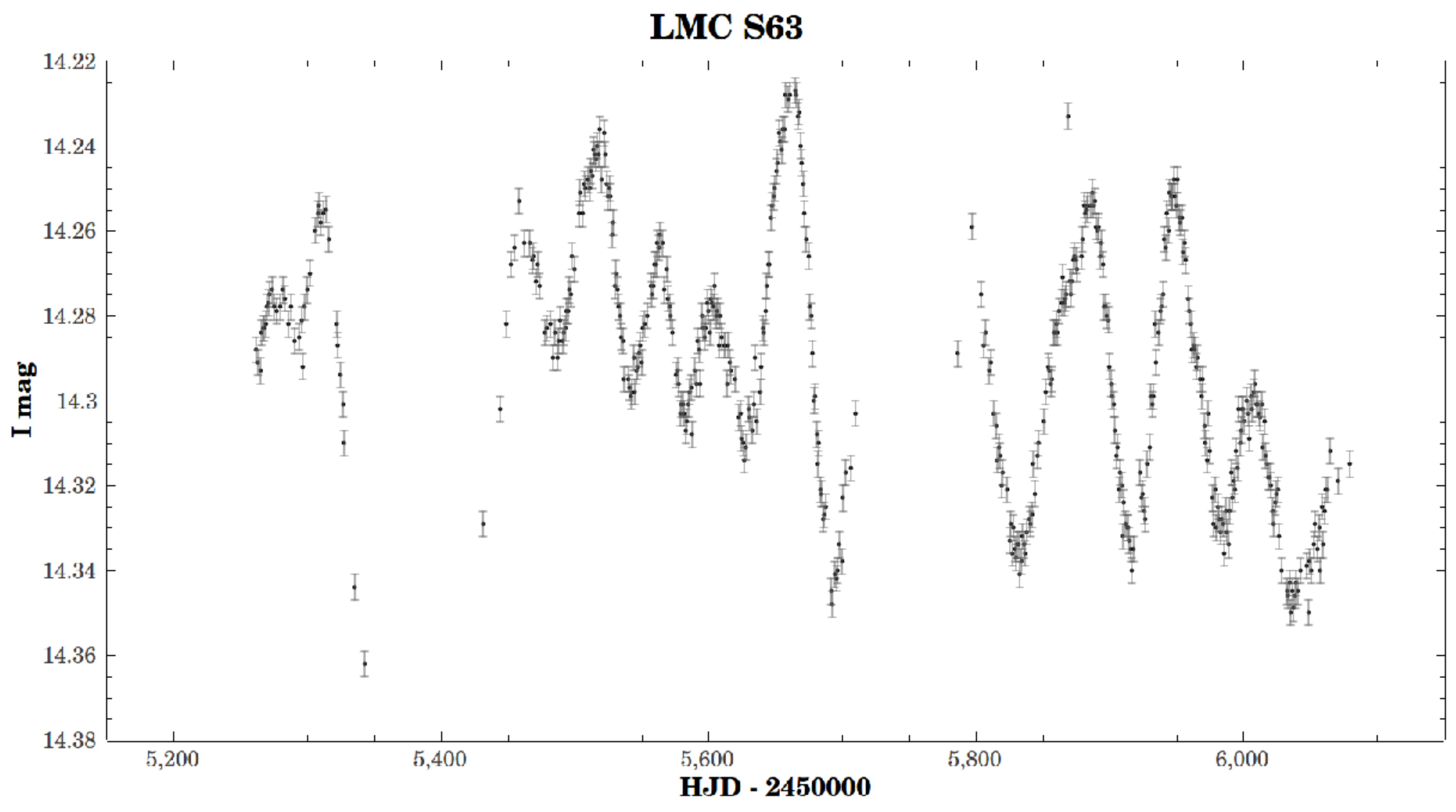}
\end{center}
\caption{OGLE-IV light curve of LMC S63.\label{fig:lmcs63_lc}}
\end{figure*}

\begin{figure*}
\begin{center}
\includegraphics[width=0.90\textwidth]{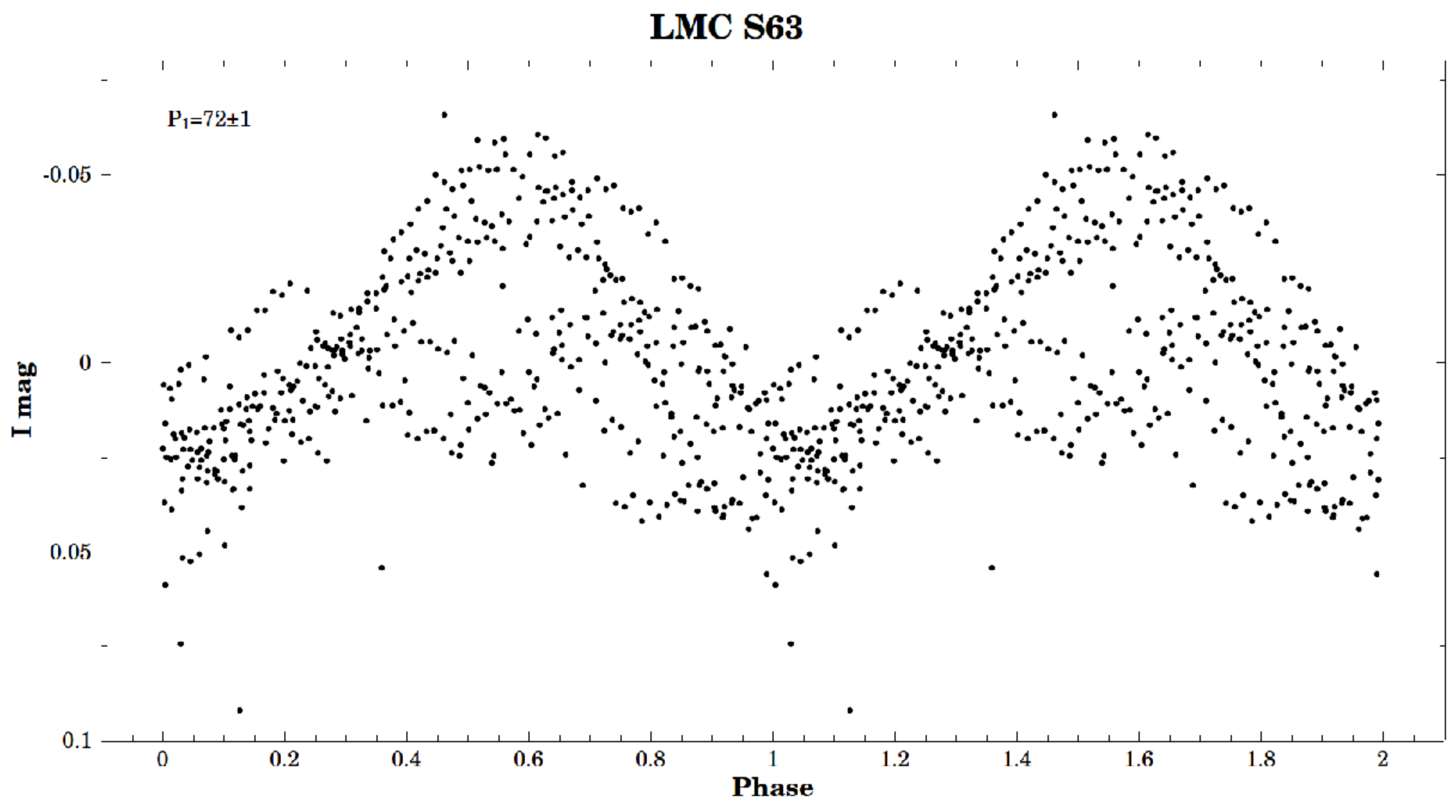}
\end{center}
\caption{The average-subtracted light curve of LMC S63 folded with the primary period P$_1$=72$\pm$1 days. \label{fig:s63_phase}}
\end{figure*}

The OGLE light curve shows a well-marked periodicity superposed on a general, slowly decreasing, trend (Fig. \ref{fig:lmcs63_lc}).  Our frequency analysis gives a primary period P$_1$=72$\pm$1 days, and a second one P$_2$=143$\pm$3 days, which is likely the P$_1$ first harmonic.  (With only 818 days in our $I$-band light curve, we are not sensitive to testing the 1260-day periodicity.) The detrended light curve of LMC S63 folded with the primary period is shown in Fig. \ref{fig:s63_phase}.  With the IR colour (Phillips 2007), the overall amplitude of the $I$-band light curve ($\Delta I$=0.135 mag), and our value of P$_1$, we classify the giant star as a LMC OSARG.  Our ephemeris is
$$HJD(min) = 2456035(\pm1) + E \times 72(\pm1)$$,
for the P$_1$ period.\\

\subsection{SMP LMC 94}
Despite having been regularly included in several surveys of Planetary Nebulae in the Magellanic Clouds, the suggestion that SMP LMC 94 could actually be a SyS dates back to the 80's when Dopita et al. (1985), although finding no trace of a red companion in the optical spectrum, observed the Raman band at $\lambda$6830 \AA. Nonetheless, in Belczy{\'n}ski et al. (2000) SMP LMC 94 is surprisingly listed as having a cool companion of spectral type M. We solve this apparent contradiction by noticing that in the reference they cite (Kontizas et al. 1996) the reported spectral type actually refers to the nearby stellar cluster, and not to SMP LMC 94 itself.  Shaw et al. (2001) observed SMP LMC 94 with HST/STIS, measuring a H$\alpha$ double-peaked velocity structure with a width of $\sim$200 km/s and a peculiar and fairly-high ionization, with prominent H$\alpha$, rather weak [OIII] lines, and broad OVI $\lambda$6830\AA. No nebular structure was resolved in the HST images.  

The SIMBAD coordinates of SMP LMC 94 mistakenly point to another star located about 15 arcsec to the SE of our target (see also Fig. \ref{fig:fc}), and this confusion has caused the target to be systematically misidentified in the literature. The correct position identifies SMP LMC 94 with 2MASS ID 05540952-7302341, as correctly reported in Phillips (2007). Its infrared colours further advocate the symbiotic character of SMP LMC 94 by placing it in the locus of genuine D-type SySs (see, e.g., Fig. 1 of Schmeja \& Kimeswenger 2003).\\

\begin{figure*}
\begin{center}
\includegraphics[width=0.90\textwidth]{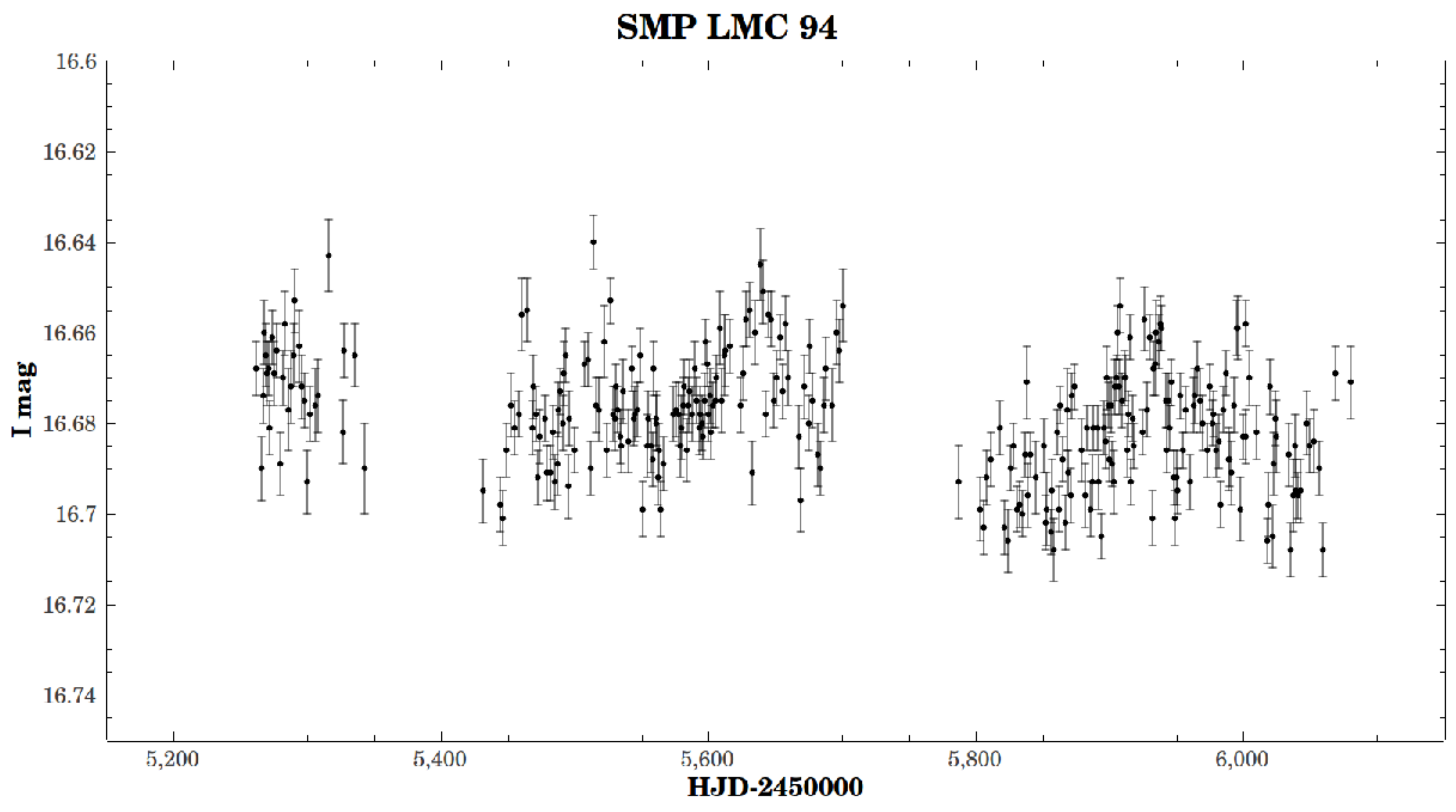}
\end{center}
\caption{OGLE-IV light curve of SMP LMC 94.\label{fig:smplmc94_lc}}
\end{figure*}

The $I$-band OGLE-IV light curve of SMP LMC 94 is presented in Fig. \ref{fig:smplmc94_lc}. SMP LMC 94 has been virtually constant around I=16.7 mag in the last two years, with a recorded maximum amplitude of only 0.07 mag. Our timing analysis finds two very weak periodicities at P$_1$=152$\pm$6 and P$_2$=30$\pm$1 days with only marginal significance.

We emphasize that the lack of a strong variability, along with the 2MASS colours that show a strong IR excess characteristic of heavily absorbed dusty SySs,
is perfectly consistent with a D-type SySs in which the dust obscuration toward the Mira variable is as strong as to completely hide the underlying pulsating photosphere.
The corresponding optical light curve will thus just reflect the variable physical conditions of the gas nebula and/or dust obscuration events -- as Kotnik-Karuza et al. (2004) demonstrated for RR Tel and as we suggest for SMP LMC 94.

\section{Discussion}
In this paper we have presented and discussed the OGLE light curves of all but one the SySs in the LMC. The analysis, performed by visual inspection, Fourier transform and other statistical methods of period searching, has allowed us to investigate for the first time in a systematic way the medium-term behaviour of these poorly studied systems. In particular, we were able in some cases to isolate and confirm the intrinsic variability of the giant component and to link it with the physical parameters of the binary system, thus collecting useful information on the character of the stellar pulsator, whose first-order pulsational ephemerides we summarize in Table \ref{tab:period}.\\

All the SySs (except for Sanduleak's star) are brighter than the tip of the Red Giant Branch in LMC at I=14.56 (Udalski 2000).  This confirms earlier results that LMC SyS giants belong to the AGB (M\"{u}rset et al. 1996; Mikolajewska 2004; Kniazev et al. 2009). \\ 

The OGLE images do not show any nebulosity around the targeted objects.  This (lack of) evidence does not imply the intrinsic absence of resolvable nebulae around SySs in the LMC, as the case of Sanduleak's star, with its almost 1 arcmin jet, spectacularly proves. \\

We have shown the importance of colour information for a self-consistent study of SySs. To optimally extract the tremendous information stored in a symbiotic light curve, the ideal case would be to perform a multi-band photometric monitoring with a frequent sampling over a well extended time baseline. The diverse phenomenology associated with SySs, dramatically reflected in their composite photometric variability, requires \textit{de facto} a multi-wavelength approach based on simultaneous multi-passband observations, to avoid any potential bias introduced by correlated flux variations (Munari et al. 1992).  Multi-band observations are essential since they provide information about the nature of the composite continuum and help identifying both the status (active or quiescent) and the emitting processes at work in a symbiotic binary. For example, very negative intrinsic colours in the blue part of the optical region, connected with optical brightening, would be the signals of energy conversion from the hot 
star to nebular emission during the development of an outburst (e.g. Angeloni et al. 2012b; Tomov et al. 2004).  The evolution in the light curve at the very beginning of an outburst would be of particular importance for mapping the process that ignites the eruption (e.g. Sokoloski et al. 2006). Furthermore, in those SySs with a high orbital inclination, the resulting eclipses would allow us to disentangle the geometry and location of the hot active region, whereas the colour indices during the totality would permit to quantify the contribution from the non-eclipsed fraction of the nebula.  A direct example of such (in our case missed) opportunity is the deeper minimum we observed in LMC N67 (Fig. \ref{fig:lmcn67_lc}) and that we speculated might be an eclipse.  With simultaneous observations in additional filter(s) we might have checked for the presence of any phase delay between the minimum at different wavelengths, delay that obviously should not be present if the variability were due to pure geometrical 
effects.\\

\begin{table*}
 \caption{Table of most relevant periods (in days) and pulsation ephemeris for the cool component in LMC SySs.}                                                                                                                                                                                                                                                                                                                                                                                                                                                                                                                                                                                                                                                                                                                                                                                                                                                                                                                                                 
                                                                                                                                                                                                                                                                                                                                                                                                                                                                                                                                
 \label{tab:period}
 \begin{tabular}{@{}cccccccccccccccc}
  \hline
   Name & P$_1$ & P$_2$ & Ephemeris\\
  \hline
 LMC S147 & 63$\pm$3 & 248$\pm$10 & Min(V) = HJD2455553 + E$\times$63\\
 LMC N19 & 506$\pm$16 & 79$\pm$1 & Min(I) = HJD2456012 + E$\times$79\\
 LMC 1 & 171$\pm$3 & 98$\pm$2 & Min(I) = HJD2455962 + E$\times$98\\
 LMC N67 & 45$\pm$5 & - & Min(I) = HJD2455513 + E$\times$45\\
 LMC S63 & 72$\pm$1 & 143$\pm$3 & Min(I) = HJD2456035 + E$\times$72\\
 SMP LMC 94 & 152$\pm$6 & 30$\pm$1 & - \\
  \hline
 \end{tabular}

\end{table*}

\section{Concluding remarks}
For future study, a dense and extended monitoring is thus the key for (1) following the ever-changing periodicities and states, (2) catching the longer periodicities, (3) promptly discovering any unpredictable sudden change in brightness due to an upcoming and/or ongoing symbiotic outburst, and (4) providing an alert for timely follow-up observations with further facilities.  This kind of monitoring is extremely time consuming and logistically demanding.  To keep following MC SySs and better characterize their properties, we have recently started an observing program (Proposal \#CN2012B-5) with the SMARTS 1.3m Telescope on Cerro Tololo in Chile. The main goal is to regularly monitor throughout the BVRI-JHK filters a selected sample of objects with a sampling of twice per week.\\  

Considering i) that the need for multi-wavelength and simultaneous observations is not important only to SyS studies, but it is also shared by many diverse science cases ranging from microlensing events to exoplanets, ii) that the optimization of telescope time use is a requirement that will be more and more crucial with the advent of the extremely large telescope era, iii) that despite being instrumentally simple, this need is not yet well satisfied by current facility instruments, we have devised an instrument concept that could represent a viable solution to all these issues. We have called this instrument BOMBOLO\footnote{The official web page of BOMBOLO is at: \texttt{http://www.aiuc.puc.cl/bombolo/}}.  As the first Chilean instrument of its kind, it is a three-armed imager covering the near-UV and optical wavelengths. The three arms work simultaneously and independently, providing synchronized imaging capability for rapid astronomical events.

The BOMBOLO Project has been recently funded through a CONICYT-QUIMAL grant (Project \#130006) and BOMBOLO has been already approved by the Board of Directors of the Southern Astrophysical Research (SOAR) 4m Telescope to become a visitor instrument. We propose it as the single solution to a series of scientific questions emerging within our science community. In particular, BOMBOLO will be able to address largely unexplored events in the minute-to-second time-scales, with the following topics for our lead science cases: 1) Simultaneous Multiband Flickering Studies of Accretion Phenomena; 2) Near UV/Optical Diagnostics of Stellar Evolutionary Phases; 3) Exoplanetary Transits; 4) Microlensing Follow‐Up; and 5) Solar System studies.  This instrument configuration will also be ideal for SySs.  Currently, capabilities for both fast and simultaneous near-UV to optical photometry are not available in the southern hemisphere.

\section*{Acknowledgment}
The authors would like to thank an anonymous referee for constructive criticism that improved the paper, and for the help in the correct identification of SMP LMC 94.
R.A. acknowledges support in the form of a PUC School of Engineering Post-doctoral Fellowship, by Proyecto GEMINI-CONICYT \#32100022, and by Proyecto Fondecyt Postdoctoral \#3100029. R.A., D.G. and T.H.P. acknowledge support by Proyecto QUIMAL \#130006. R.A., C.E.F.L., C.N. and M.C. acknowledge support by the Chilean Ministry for the Economy, Development, and Tourism's Programa Iniciativa Cient\'{i}fica Milenio through grant P07-021-F, awarded to The Milky Way Millennium Nucleus; by the BASAL Center for Astrophysics and Associated Technologies (PFB-06); and by Proyecto Anillo ACT-86. M.C., C.N., and C.E.F.L. also acknowledge support by Proyecto Fondecyt Regular \#1110326. C.E.F.L. acknowledges a Post-Doctoral fellowship of the CNPq 150632/2013-4 and of the INCT-INEspa\c{c}o. C.N also acknowledges support from CONICYT-PCHA/Mag{\'i}ster Nacional/2012-22121934 and Fondecyt Regular \#1110326. N.M. thanks the Departamento de Astronom\'{i}a y Astrof\'{i}sica of the Pontificia Universidad Cat\'olica de Chile in 
Santiago for the warm hospitality during the preparation of this paper. P.P. is supported by the INTEGRAL ASI-INAF grant \#033/10/0. D.G. acknowledges financial support from ALMA-CONICYT grants \#31080006, \#31120017 and FONDECYT \#11110149. The OGLE project has received funding from the European Research Council under the European Community's Seventh Framework Programme (FP7/2007-2013)/ ERC grant agreement \#246678. This paper utilizes public domain data obtained by the MACHO Project, jointly funded by the US Department of Energy through the University of California, Lawrence Livermore National Laboratory under contract No. W-7405-Eng-48, by the National Science Foundation through the Center for Particle Astrophysics of the University of California under cooperative agreement AST-8809616, and by the Mount Stromlo and Siding Spring Observatory, part of the Australian National University. This research has also made use of the ASI Science Data Center Multimission Archive. We would like to thank the Harvard 
College Observatory for its hospitality and access to their archival plate 
collection. 

{}

\end{document}